\documentclass{article}
\usepackage{amsfonts}
\usepackage{eurosym}
\usepackage{amsmath,graphicx}
\usepackage[margin=1in]{geometry}
\usepackage{amssymb,amsmath,amsthm,mathtools}
\usepackage{mathrsfs}
\usepackage{hyperref}
\usepackage{subcaption}
\usepackage{verbatim}
\usepackage{mdwlist}
\usepackage{bm}
\usepackage{geometry}
\usepackage{mathrsfs}
\usepackage{enumerate}
\usepackage{color}
\usepackage{subcaption}
\usepackage{epsfig,color}

\begin{document}

\title{A confined rod: mean field theory for hard rod-like particles}
\author{ Jamie M. Taylor $^1$\thanks{%
0000-0002-5423-828X, jamie.taylor@cunef.edu}, Thomas G. Fai $^2$\thanks{%
0000-0003-0383-5217, tfai@brandeis.edu}, Epifanio G. Virga$^3$\thanks{%
0000-0002-2295-8055, eg.virga@unipv.it}, \\
Xiaoyu Zheng $^{4}$\thanks{%
0000-0002-3787-7741, xzheng3@kent.edu}, Peter Palffy-Muhoray $^{4,5}$\thanks{%
0000-0002-9685-5489, mpalffy@kent.edu}\\
\emph{$^{1}$ CUNEF Universidad, Madrid, Spain}\\
\emph{$^{2}$ Department of Mathematics and Volen Center for Complex Systems,}%
\\
\emph{\ Brandeis University Waltham, MA, USA} \\
\emph{$^3$ Dipartimento di Matematica, Universit\`{a} di Pavia, Pavia, Italy 
}\\
\emph{$^{4}$ Department of Mathematical Sciences, Kent State University,
Kent, \ \ OH, USA }\\
\emph{$^{5}$Advanced Materials and Liquid Crystal Institute, Kent State
University, Kent, OH, USA} }
\maketitle

\begin{abstract}
In this paper, we model the configurations of a system of hard rods by
viewing each rod in a cell formed by its neighbors. By minimizing the free
energy in the model and performing molecular dynamics, where, in both cases,
the shape of the cell is a free parameter, we obtain the equilibrium
orientational order parameter, free energy and pressure of the system. Our
model enables the calculation of anisotropic stresses exerted on the walls
of the cell due to shape change of the rod in photoisomerization. These
results are a key step towards understanding molecular shape change effects
in photomechanical systems under illumination.
\end{abstract}

\section{Introduction}

Our motivation for this work is to gain insights into photomechanical
materials which convert light energy directly into mechanical work. Azo-dye
containing nematic liquid crystal elastomers (LCEs) are one example of such
materials \cite{swimming, white, Kuzyk, guo}. Under illumination, dye
molecules can undergo photoisomerization, changing their shape, resulting in
the macroscopic deformation of samples, which can exert forces and do
mechanical work.

Here we aim to understand the mechanics of bulk shape change of such
materials from a microscopic perspective. Specifically, we consider a system
of hard rod-like particles which interact via hard-core steric interactions,
conserving linear and angular momentum and energy. At low number densities,
we expect the system to be isotropic, but as the number density is
increased, a transition to an orientationally ordered phase is expected.
Unlike in Onsager theory \cite{Onsager1949}, where the pair-excluded volume
plays an essential role, we regard each rod as being effectively confined in
a local cell -- a rectangular box or an ellipsoid -- created by neighboring
particles, where the free volume of each particle in its cell determines the
phase behavior.

There are various cell models for hard spheres \cite{Devonshire1937,
Eyring1941}. In cell theories, the free volume for non-interacting rigid
spheres is defined as that volume of the cell in which the center of mass of
a particular molecule is able to move when all of the other molecules in the
liquid or dense gas are held fixed at their mean lattice positions. The
single occupancy model, with one particle per cell, works well at high
densities, but less well at low densities. To rectify the situation, the
concept of `communal entropy' was introduced to describe the solid-fluid
phase transition \cite{Kirkwood1950}.

As a first attempt for rod-like particles, in this paper, we shall only
consider the single occupancy model where each rod is confined to its
private cell. Rather than considering the precise geometry of the boundary
of the cell described by the neighboring particles, we assume that the cell
can be approximated via a \textit{geometric} mean field, formed from a
simple geometry (cuboidal or ellipsoidal) of given volume. We anticipate our
results to produce quantitatively accurate thermodynamic properties, such as
orientational order, pressure and energy in the high density regime.

The paper is organized as follows. In section \ref{Sec_theory}, we lay out
our mean-field theory, where each particle is regarded as being confined to
a cell defined by its neighbors. The free volume of a particle in a given
cell depends on its orientation; the free energy depends on the average free
volume over all possible particle orientations available to the center of
the particle in its cell. At a given volume fraction, the cell adopts a
shape which maximizes the average free volume, or, equivalently, minimizes
the free energy. In section \ref{Sec_ellip_Rec}, we apply the theory to the
case of an ellipsoid in a rectangular cell, and compare the results with
those from molecular dynamics simulations. We also consider the limiting
case of a needle in an ellipsoidal cell. In section \ref{Sec_Disscussion},
we discuss the effects of shape change in both rectangular and ellipsoidal
cells. The paper is summarized in the conclusion section \ref{Sec_Conclusion}%
.

\section{Mean-field theory}

\label{Sec_theory}

The Helmholtz free energy for a system of $N$ indistinguishable particles in
volume $V$ at temperature $T$ is given by \cite{Eyring1941} 
\begin{equation}
F=-kT\ln Z_{tot},  \label{eq_energy_1}
\end{equation}%
where $k$ is Boltzmann's constant and $Z_{tot}$ is the total partition
function. After integrating over momenta, this is given by 
\begin{equation}
Z_{tot}=\frac{1}{\Lambda _{dB}^{3N}N!}\int e^{-\frac{H(\mathbf{q}_{1},...,%
\mathbf{q}_{N})}{kT}}d\mathbf{q}_{1}...d\mathbf{q}_{N},  \label{eq_Z_tot}
\end{equation}%
where $\Lambda _{dB}$ is the de Broglie wavelength, $\mathbf{q}_{i}=(\mathbf{%
r}_{i}, \mathbf{\hat{m}}_{i})$ is the generalized coordinate of the $i-$th
particle, where $\mathbf{r}_{i}$ is the position of center of mass and $%
\mathbf{\hat{m}}_{i}$ is a unit vector along the symmetry axis of the
particle.

The essence of mean field theory is to approximate the potential energy of
the system by the sum of identical single-particle potentials, which
describe the interaction of each particle with an effective mean field \cite%
{Kuzemsky}. Single-particle potentials are not unique, but they must satisfy
certain consistency conditions \cite{PPM}. In this work, we assume that
there are no long-range, attractive interactions and the particles are kept
in a prescribed volume $V$ by an adjustable external pressure. Since the
interactions are steric, the particles cannot overlap. As a first attempt,
then, following the single-particle potential construction procedure in \cite%
{PPM}, we obtain the single particle potential,%
\begin{equation}
U(\mathbf{q)=}\left\{ 
\begin{array}{cc}
0, & \text{if particle is wholly within the cell, and} \\ 
\infty & \text{if it is not,}%
\end{array}%
\right.  \label{SP}
\end{equation}%
which satisfies the consistency requirements. Each particle is thus confined
to a cell $\mathcal{C}$ defined by the surrounding particles. Although in a
real system the cells have non-uniform distributions and fluctuate in time,
here, in order to make the model tractable, we assume that the cells are
frozen in time and all of equal size.

The total partition function is 
\begin{equation}
Z_{tot}=z^{N}=\left( \frac{1}{\Lambda _{dB}^{3}}\int e^{-\frac{U(\mathbf{q}%
_{1})}{kT}}d\mathbf{q}_{1}\right) ^{N},  \label{eq_Z_tot_2}
\end{equation}%
where $z$ is the single-particle partition function, and $U(\mathbf{q}_{1})$
is the single-particle potential in Eq.~\eqref{SP}. Being now in individual
cells, all particles are effectively distinguishable and there is no need
for the Boltzmann factor $1/N!$ as in Eq.~\eqref{eq_Z_tot}.

The integration over position $\mathbf{r}_{1}$ in Eq.~\eqref{eq_Z_tot_2}
gives the free volume $V_{f}(\mathbf{\hat{m},}\mathcal{C})$, the volume
accessible to the center of the particle in the cell $\mathcal{C}$ for a
given particle orientation $\mathbf{\hat{m}}$. That is, 
\begin{equation}
z=\frac{1}{\Lambda _{dB}^{3}}\int e^{-\frac{U(\mathbf{q}_{1})}{kT}}d\mathbf{q%
}_{1}=\frac{1}{\Lambda _{dB}^{3}}\int_{S^{2}}\int_{\mathcal{C}}e^{-\frac{U(%
\mathbf{q}_{1})}{kT}}d\mathbf{r}_{1}d\mathbf{m}_{1}=\frac{1}{\Lambda
_{dB}^{3}}\int_{S^{2}}V_{f}(\mathbf{\hat{m}}_{1},\mathcal{C})d\mathbf{\hat{m}%
}_{1}.
\end{equation}
The total partition function in Eq.~\eqref{eq_Z_tot_2} can then be rewritten
as%
\begin{equation}
Z_{tot}(\mathcal{C)}=\left( \frac{1}{\Lambda _{dB}^{3}}\int_{S^{2}}V_{f}(%
\mathbf{\hat{m},}\mathcal{C})d\mathbf{\hat{m}}\right) ^{N}.
\label{eq_Z_tot_3}
\end{equation}%
The free energy in Eq.~\eqref{eq_energy_1} can be approximated, to within an
inessential additive constant, as 
\begin{equation}
F=-NkT\ln \frac{1}{\Lambda _{dB}^{3}}\int_{S^{2}}V_{f}(\mathbf{\hat{m},}%
\mathcal{C})d\mathbf{\hat{m}}.  \label{fren}
\end{equation}%
The free energy density $\mathcal{F}=F/(NV_{cell})$ may be written in the
more familiar form 
\begin{equation}
\mathcal{F}=kT\rho _{0}\ln \rho _{0}-kT\rho _{0}\ln \frac{1}{\Lambda
_{dB}^{3}}\frac{1}{V_{cell}}\int_{S^{2}}V_{f}(\mathbf{\hat{m}},\mathcal{C})d%
\mathbf{\hat{m}},
\end{equation}%
where $\rho _{0}=1/V_{cell}$ is the number density, but for our purposes Eq.~%
\eqref{fren} suffices.

The orientational probability density function $\rho (\mathbf{\hat{m})}$ is 
\begin{equation}
\rho (\mathbf{\hat{m}})=\frac{V_{f}(\mathbf{\hat{m}},\mathcal{C})}{%
\int_{S^{2}}V_{f}(\mathbf{\hat{m}},\mathcal{C})d\mathbf{\hat{m}}},
\end{equation}%
and the orientational order parameter tensor $\mathbf{Q}$ is%
\begin{equation}
\mathbf{Q}=\frac{\int_{S^{2}}\frac{1}{2}(3\mathbf{\hat{m}\hat{m}}-\mathbb{I}%
)V_{f}(\mathbf{\hat{m},}\mathcal{C})d\mathbf{\hat{m}}}{\int_{S^{2}}V_{f}(%
\mathbf{\hat{m},}\mathcal{C})d\mathbf{\hat{m}}},
\end{equation}%
where is $\mathbb{I}$ the identity tensor. The scalar order parameters are
projections of $\mathbf{Q}$ onto the principal directions of the cell $%
\mathbf{\hat{e}}_{i}\mathbf{\hat{e}}_{i}$, and 
\begin{equation}
S_{2i}=\left\langle \mathcal{P}_{2}(\mathbf{\hat{m}\cdot \hat{e}}%
_{i})\right\rangle ,
\end{equation}%
where $\mathcal{P}_{2}(x)$ is the second Legendre polynomial. For
simiplicity, we drop the subscript $i$, and understand that $\mathbf{\hat{e}}
$ refers to the distinguished direction of the cell. The order parameter
tensor is the normalized traceless second moment of the orientational
distributions function; the first moment vanishes due to the quadrupolar
symmetry of the rodlike particles. We focus primarily on $\mathbf{Q}$ in
this work, however, higher order moments are also present, and shall be
discussed below.

In general, self-consistency requires that free parameters of the mean-field
pseudopotential be chosen so as to minimize the free energy. In the present
setting, it is the cell shape itself that needs to be so chosen. We next
address this issue in general for a given cell volume $V_{cell}$ before
considering more specific instances.

\subsection{Cell shape and pressure}

We first show that, independent of the shape of the cell, the Cauchy stress
tensor $\boldsymbol{\sigma}$, associated at equilibrium with the free energy 
$F$, is isotropic, and so its effect on the walls of the cell reduces to
uniform pressure. Here we consider the minimiser $\mathcal{C}$ over a family
of cells described as a linear, volume preserving transformation of a
reference cell $\mathcal{C}_0$. We assume that the cell $\mathcal{C}_{0}$ is
of some arbitrary shape (say, a parallelepiped, for definiteness), and let $%
\mathbf{F}$ denote a corresponding deformation gradient. Thus $F$ can be
regarded as a function of $\mathbf{F}$, whose explicit form need not be
specified at this stage, provided that it is sufficiently smooth. The Cauchy
stress $\boldsymbol{\sigma }$ exerted on the walls of $\mathcal{C}$ is \cite{Spencer}
\begin{equation}
\boldsymbol{\sigma }=\frac{1}{\det {\mathbf{F}}}\frac{\partial F}{\partial {%
\mathbf{F}}}\cdot {\mathbf{F}}^{T}.  \label{eq_var_1}
\end{equation}

For $\mathcal{C}$ and $\mathcal{C}_0$ to have the same volume $V_{cell}$, $%
\mathbf{F}$ must obey the constraint 
\begin{equation}
\det\mathbf{F}\equiv 1.
\end{equation}
Stationarity of $F$ over this constraint requires that 
\begin{equation}
\frac{\partial F}{\partial \mathbf{F}} = \lambda \frac{\partial}{\partial 
\mathbf{F}}\det\mathbf{F},  \label{eq_var_2}
\end{equation}
where $\lambda$ is a Lagrange multiplier. Since, for an invertible tensor $%
\mathbf{F}$, 
\begin{equation}
\frac{\partial}{\partial \mathbf{F}}\det{\mathbf{F}}=(\det\mathbf{F}) 
\mathbf{F}^{-T},  \label{eq_var_3}
\end{equation}
it follows from Eqs.~\eqref{eq_var_1}, \eqref{eq_var_2}, and \eqref{eq_var_3}
that

\begin{equation}
\boldsymbol{\sigma }=\frac{1}{\det {\mathbf{F}}}\frac{\partial F}{\partial {%
\mathbf{F}}}\cdot {\mathbf{F}}^{T}=\lambda \mathbb{I}.  \label{eq_iso_pres}
\end{equation}%
Eq.~\eqref{eq_iso_pres} indicates that the pressure is isotropic in
equilibrium, and the Lagrange multiplier reduces to the (isotropic) pressure.

We remark that a stress also acts on the particle inside the cell, and if
the pressure is isotropic on the cell walls, it will not, in general, be
isotropic on the particle - and vice versa.

\section{Results}

\label{Sec_ellip_Rec}

In this section, we first present results for a uniaxial ellipsoid in a
rectangular cell where the free volume is calculated by numerical
integration, and then compare these results with those from molecular
dynamics. We then present the case of a needle in an ellipsoidal cell, where
an analytical solution is possible.

\subsection{ Uniaxial ellipsoid in a rectangular cell}

\label{Sec_free_vol}

Here we consider the case when the cell is a parallelepiped and the particle
is a uniaxial ellipsoid. We conjecture, as suggested by numerics, that the
orientational averaged free volume of an ellipsoid in a parallelepiped
cannot attain its maximum and free energy its minimum when the cell faces
are not orthogonal. Furthermore, if the particle has uniaxial symmetry, the
maximum free volume occurs when the cell is uniaxial. Therefore here we
consider uniaxial rectangular cells, and apply the mean field theory of
Section \ref{Sec_theory} to the case of a rigid uniaxial ellipsoid in a
uniaxial rectangular cell.

Let $a$ and $b$ be the lengths of the semiaxes of the uniaxial ellipsoid in
the directions along and perpendicular to the symmetry axis. The equation of
the ellipsoid is 
\begin{equation}
\mathbf{rAr}=1,
\end{equation}%
where%
\begin{equation}
\mathbf{A}=\frac{1}{b^{2}}\mathbb{I}+(\frac{1}{a^{2}}-\frac{1}{b^{2}})%
\mathbf{\hat{m}\hat{m},}
\end{equation}%
and $\mathbf{\hat{m}}$ is along the symmetry axis. The aspect ratio, the
ratio of the length of the distinguished semi-axis to that of one of the
others, is $\ \eta =a/b$. Alternately, we can consider the uniaxial affine
deformation of a sphere of radius $R$ to form this ellipsoid; the principal
stretches are $\lambda _{1}=a/R$, $\lambda _{2}=\lambda _{3}=b/R$. If the
volume is conserved, $\lambda _{2}=\lambda _{3}=1/\sqrt{\lambda _{1}}$, and
the aspect ratio 
\begin{equation}
\eta =\lambda ^{3/2},
\end{equation}%
where $\lambda$ indicates stretch along the distinguished principal
direction. The aspect ratio $\eta$ and stretch $\lambda $ will be used
interchangeably to describe particle and cell shapes.

The particle volume is $v_{p}=4\pi ab^{2}/3$ and the occupied volume
fraction is $\phi =v_{p}/V_{cell}=v_{p}\rho _{0}$.

\begin{figure}[h]
\begin{center}
\includegraphics[width=0.4\textwidth]{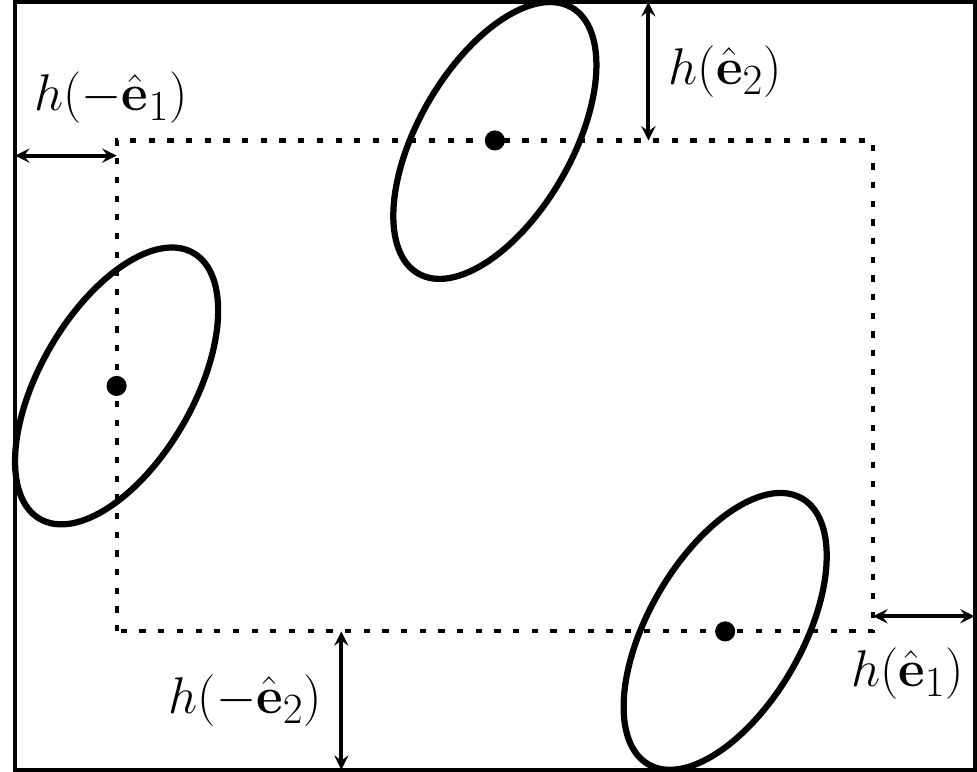} %
\end{center}
\caption{Illustration of the free volume available to a convex particle in a
polygonal cavity.}
\label{fig_free_volume}
\end{figure}

The extent of the ellipsoid with orientation $\mathbf{\hat{m}}$ in direction 
$\mathbf{\hat{N}}$ is%
\begin{equation}
h(\mathbf{\hat{N})}=\sqrt{a^{2}(\mathbf{\hat{m}\cdot \hat{N}})^{2}+b^{2}(1-(%
\mathbf{\hat{m}\cdot \hat{N}})^{2})},
\end{equation}%
and in the principal directions of the cell this is 
\begin{equation}
h_{i}=\sqrt{a^{2}m_{i}^{2}+b^{2}(1-m_{i}^{2})},i=1,2,3.
\end{equation}%
The free volume available to the center of the ellipsoid is then simply 
\begin{equation}
V_{f}(\mathbf{\hat{m}},\mathcal{C})=\prod_{i=1}^{3}\max \{d_{i}-2h_{i},0\},
\label{eq_V_free}
\end{equation}%
where $d_i$'s are lengths of the cell along its principal directions. A 2D
illustration of the free volume of an ellipse in a rectangular cell is given
in Fig.~\ref{fig_free_volume}. 

The free energy of one cell, from Eq.~\eqref{fren}, is 
\begin{equation}
F_{c}=-kT\ln \int_{S^{2}}\prod_{i=1}^{3}\max (d_{i}-2h_{i}(\mathbf{\hat{m}}%
),0)d\mathbf{\hat{m}},  \label{nin}
\end{equation}%
where we have omitted $-3\ln \Lambda _{dB}$. For simplicity, we will drop
the subscript, and with the understanding that the free energy $F$ is for a
single cell. The free energy $F$ can therefore be calculated by
straightforward numerical integration of Eq.~\eqref{nin} using, say, the
quasi-Monte Carlo method. Local minima/maxima can be identified via standard
gradient descent/ascent methods.

Let $d_{2}=d_{3}$ be the lengths of the cell along two principal directions,
and $d_{1}$ be the length along the third, the distinguished principal
direction. The aspect ratio of the cell is $\eta_{cell} =d_{1}/d_{2}$; the
stretch is $\lambda_{cell} =\eta_{cell}^{2/3}$. The cell is considered
prolate if $\eta_{cell} >1$ and oblate if $\eta_{cell} <1$. In calculating
the equilibrium free energy, at each volume fraction, we use $\eta_{cell}$,
which minimizes the free energy, to satisfy self-consistency.

We note that if the maximum (linear) dimension of the ellipsoid exceeds the
minimum dimension of the cell, then the cell will exclude some orientations
of the ellipsoid. In the case of a cubic domain, with $\eta_{cell} = 1$,
this occurs at a critical volume fraction; the corresponding critical volume
fractions are 
\begin{equation}
\phi _{cp}=\frac{\frac{4\pi }{3}ab^{2}}{(2a)^{3}}=\frac{\pi }{6}\left( \frac{%
b}{a}\right) ^{2}
\end{equation}%
for prolate and%
\begin{equation}
\phi _{co}=\frac{\frac{4\pi }{3}ab^{2}}{(2b)^{3}}=\frac{\pi }{6}\left( \frac{%
a}{b}\right)
\end{equation}%
for oblate ellipsoids.

\begin{figure}[h]
	\centering
\begin{subfigure}[c]{0.45\linewidth}
	\centering
	\includegraphics[width=\linewidth]{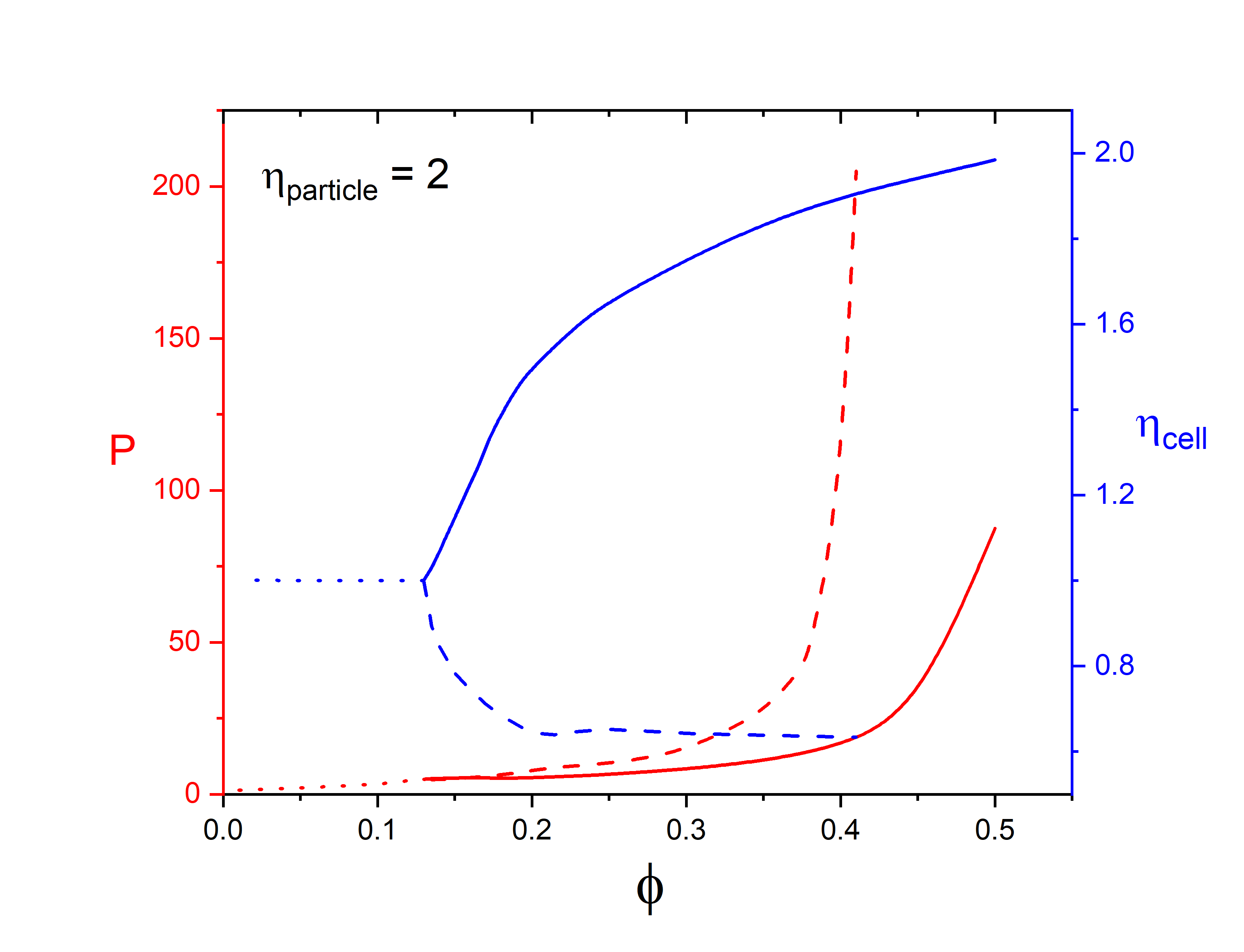}
	\caption{}
\end{subfigure}
\begin{subfigure}[c]{0.45\linewidth}
	\includegraphics[width=\linewidth]{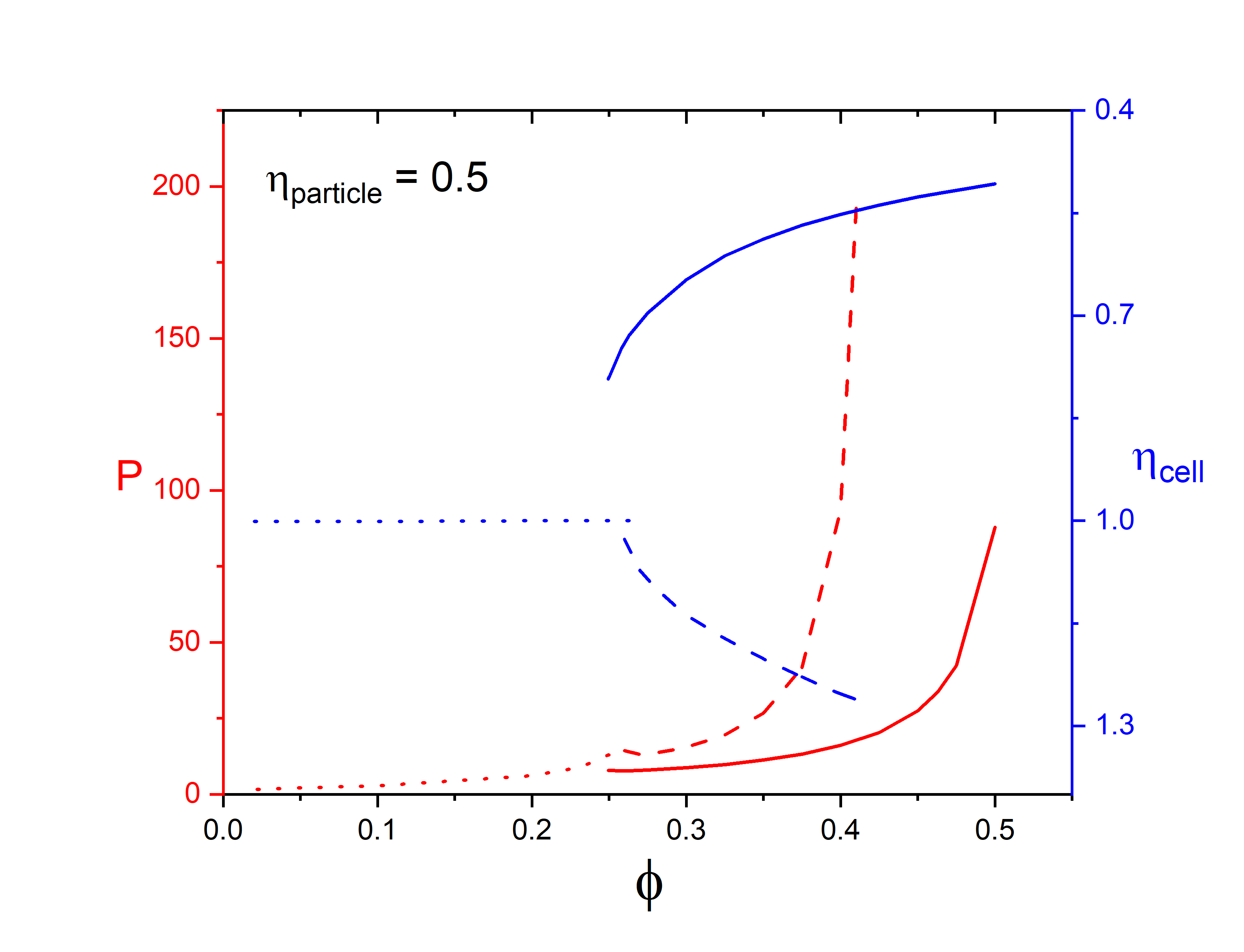}
	\caption{}
\end{subfigure}
\begin{subfigure}[c]{0.45\linewidth}
	\centering
	\includegraphics[width=\linewidth]{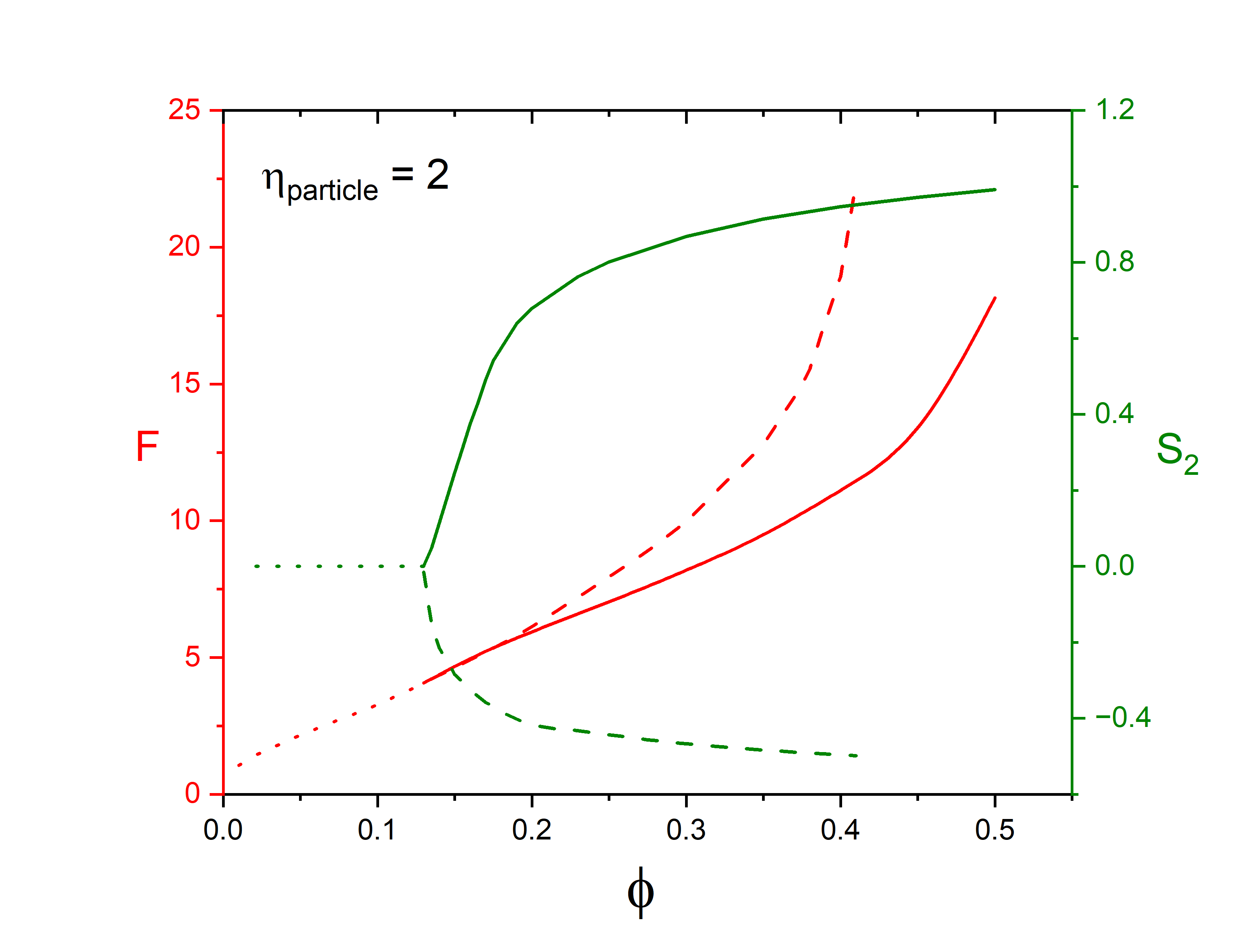}
	\caption{} 
\end{subfigure}
\begin{subfigure}[c]{0.45\linewidth}
	\centering
	\includegraphics[width=\linewidth]{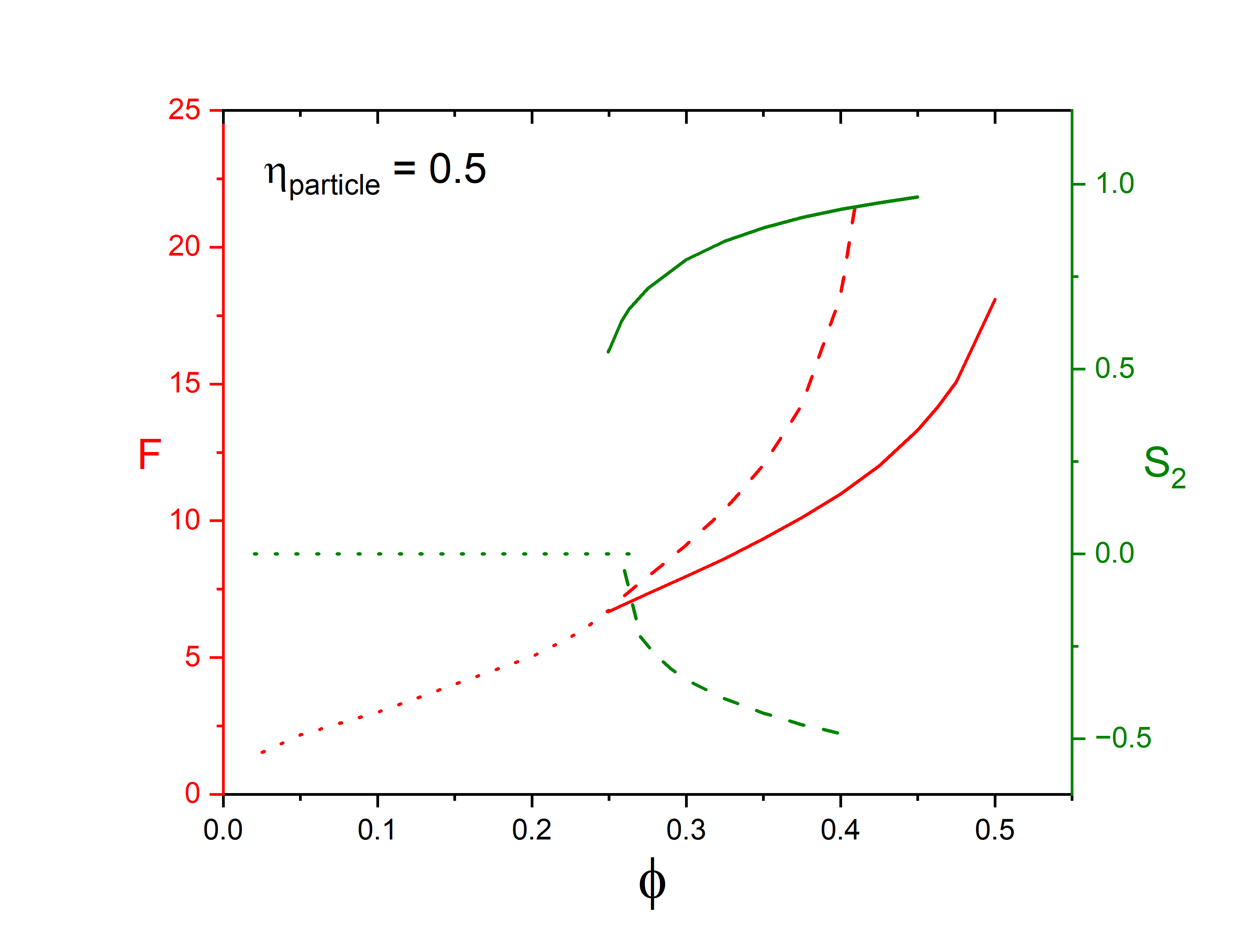}
	\caption{}
\end{subfigure}
\caption{Results for uniaxial ellipsoids with aspect ratio $\protect\eta%
_{particle}=2$ (left column) and $\protect\eta_{particle}=0.5$ (right
column) in a rectangular box. Top row: pressure $P$ (left axes) and cell aspect ratio $\eta_{cell}$ (right axes) vs.~volume fraction $\phi$. Bottom row: free energy $F$ (left axes) and scalar order parameter $S_2$ (right axes) vs.~volume fraction $\phi$.  Only locally stable solutions are shown. Dotted curves are for
isotropic phases where the cell is a cube, the dashed curves are for oblate
phases with $S_2<0$ and solid curves are for prolate phase with $S_2>0$.}
\label{fig:bif}
\end{figure}

\begin{figure}[h]
	\centering
\begin{subfigure}[c]{0.25\linewidth}
		\centering
		\includegraphics[width=\linewidth]{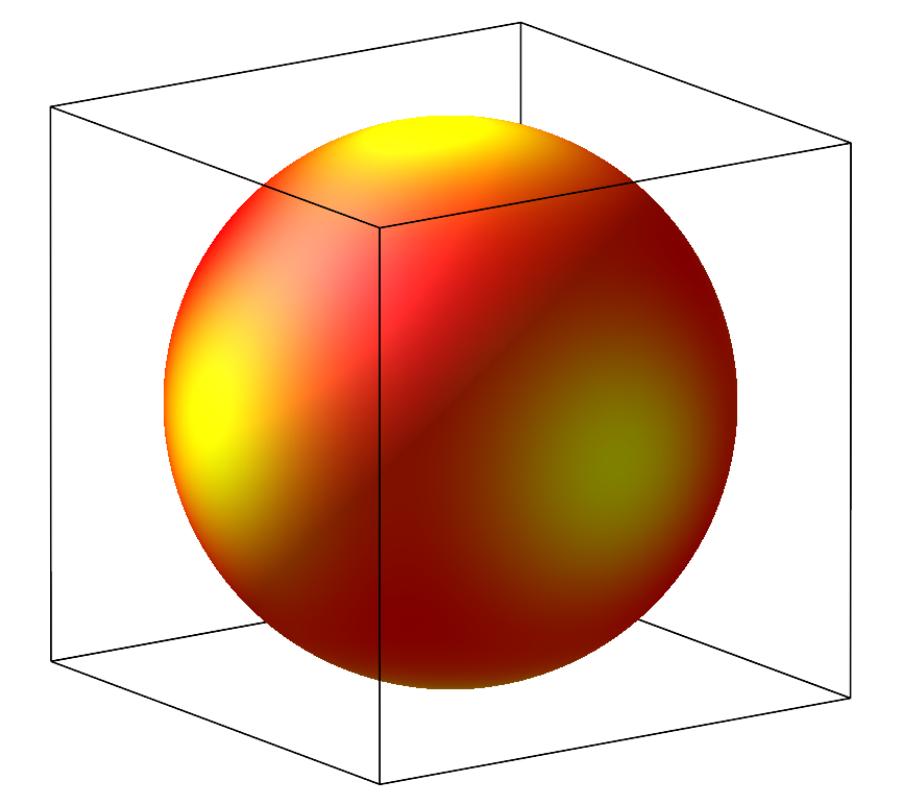}\caption{$(\phi, \eta_{cell}) = (0.13, 1) $}
\end{subfigure}\qquad
\begin{subfigure}[c]{0.25\linewidth}
		\centering
		\includegraphics[width=\linewidth]{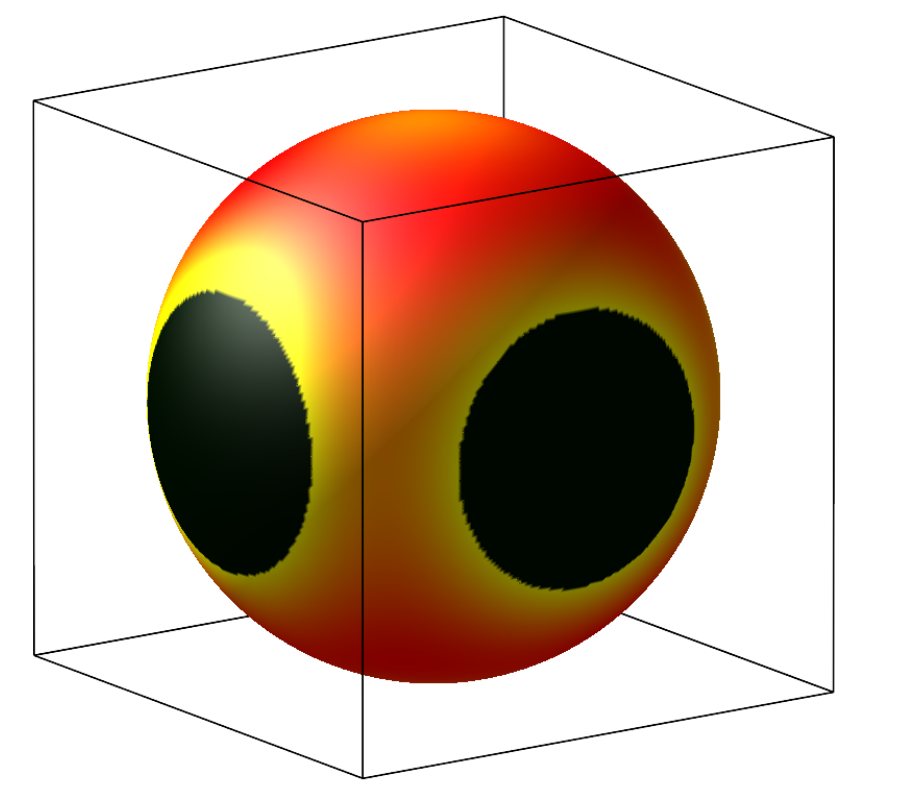}\caption{$(\phi, \eta_{cell}) = (0.153, 1.17) $}
	\end{subfigure}\qquad
\begin{subfigure}[c]{0.25\linewidth}
		\centering
		\includegraphics[width=\linewidth]{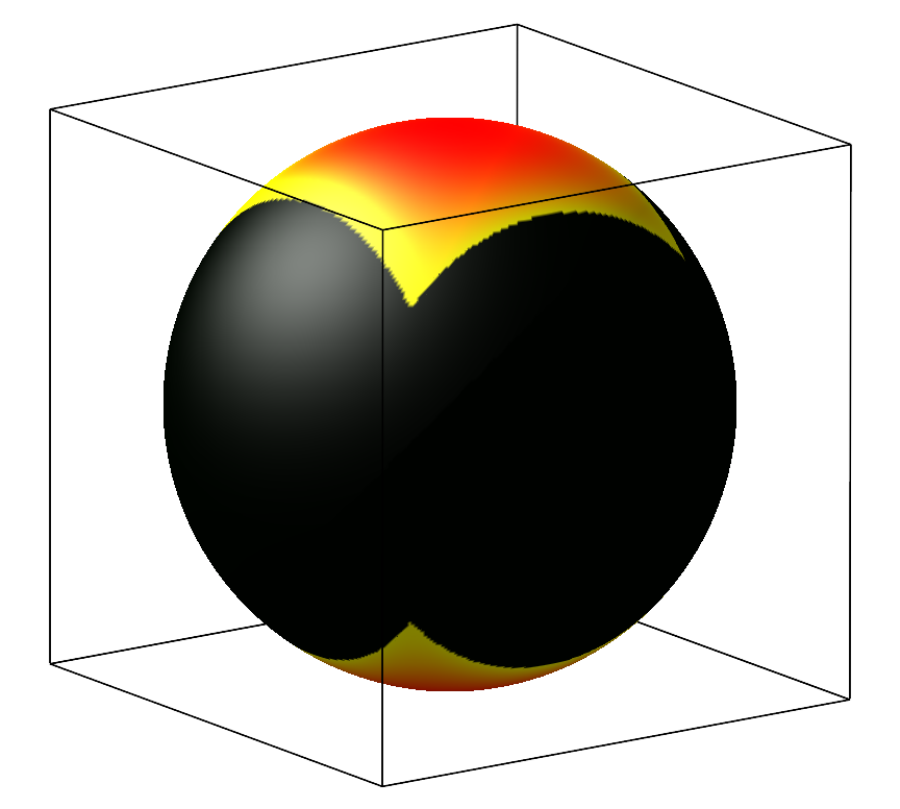}
		\caption{$(\phi, \eta_{cell}) = (0.23, 1.57)$} 
\end{subfigure}
\begin{subfigure}[c]{0.25\linewidth}
		\centering
		\includegraphics[width=\linewidth]{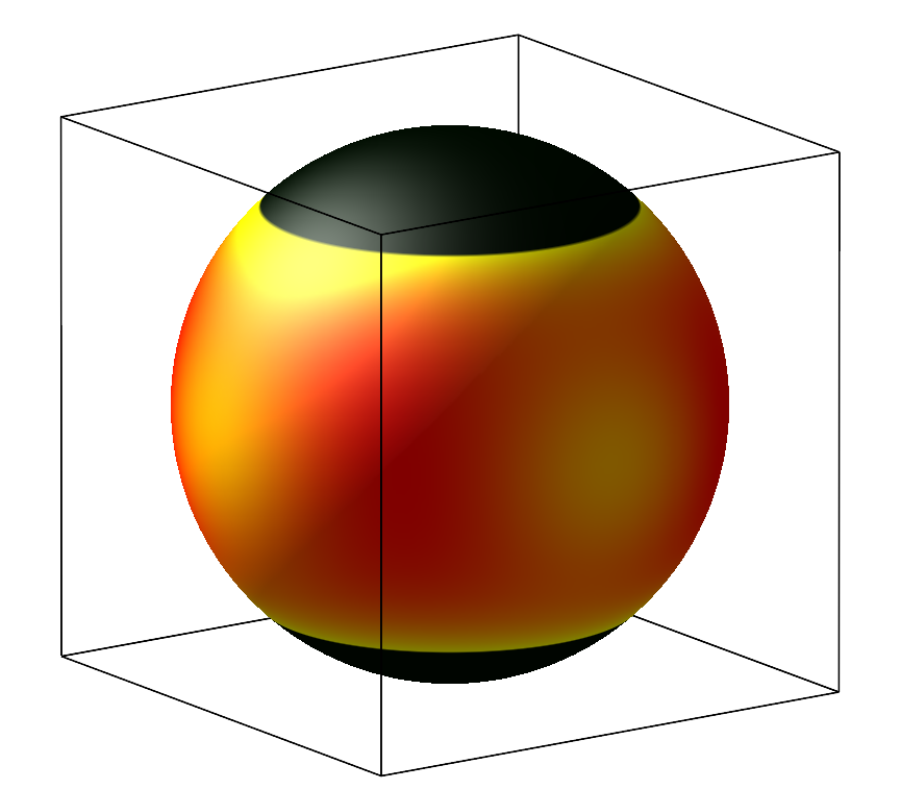}
		\caption{$(\phi, \eta_{cell}) =  (0.15, 0.78) $ }
\end{subfigure}\qquad
\begin{subfigure}[c]{0.25\linewidth}
		\centering
		\includegraphics[width=\linewidth]{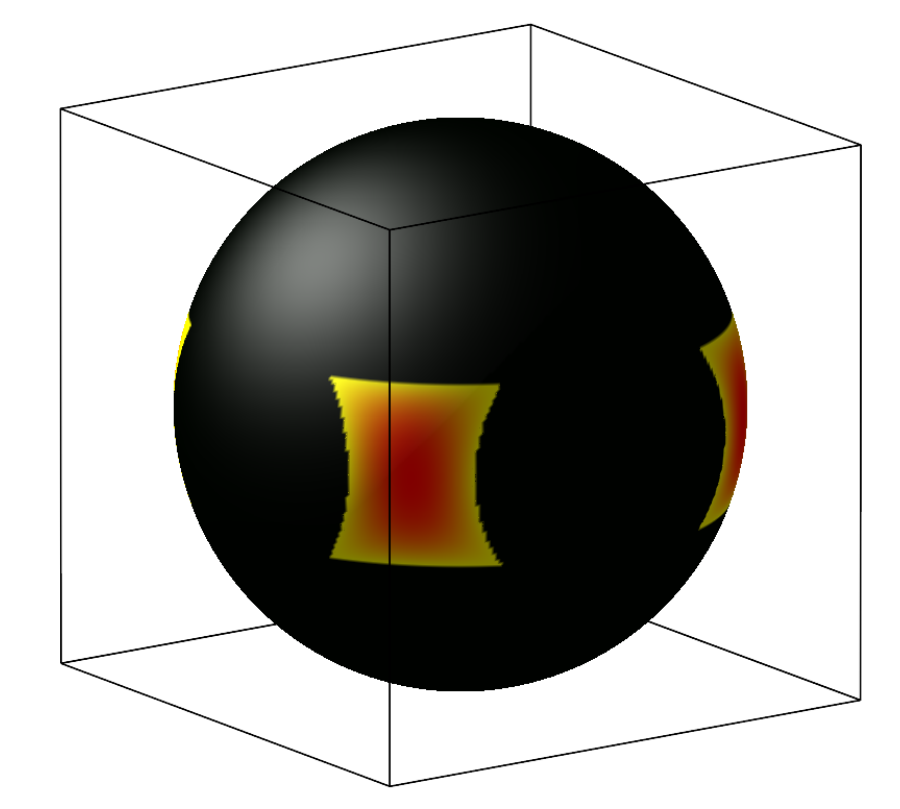}\caption{$(\phi, \eta_{cell}) = (0.29, 0.65)$ }
\end{subfigure}\qquad
\begin{subfigure}[c]{0.25\linewidth}
	\centering
	\includegraphics[width=0.15\linewidth]{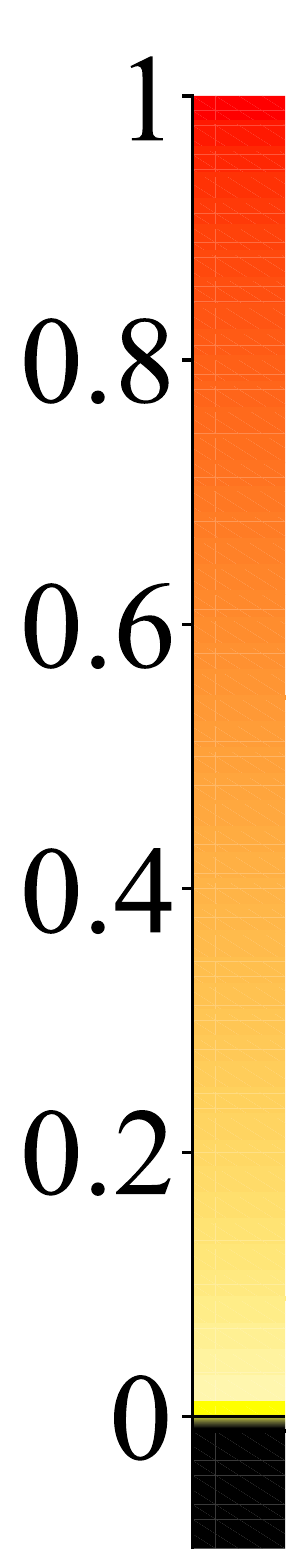}
\end{subfigure}
\caption{Orientational distribution function for a uniaxial prolate
ellipsoid with aspect ratio $\protect\eta_{particle}=2$ at various volume
fractions $\protect\phi$ in a cube $\eta_{cell}=1$(a), a prolate cell with $\protect\eta_{cell}>1$ (b,c) and an oblate cell with $%
\protect\eta_{cell}<1$ (d,e). Each plot is scaled so that its maximum value
is 1, and the regions in black correspond to where the probability is zero. }
\label{fig:pdf}
\end{figure}

Figure \ref{fig:bif} shows the results for two representative ellipsoids
with aspect ratios $2$ and $1/2$, respectively, as function of occupied
volume fraction. Volume fraction was varied by changing particle volumes at
fixed particle aspect ratio. Anisotropy at large volume fractions is
apparent both in the cell shape, and in the orientational distribution
function. Four scenarios have been observed: prolate or oblate particles in
prolate or oblate cells. If the particle and cell shape are similar, then $%
S_{2}>0$. If the particle and cell shapes are dissimilar, then $S_{2}<0$.

For the prolate ellipsoid with $\eta_{particle}=2$, the system is dilute
with $\phi <\phi _{cp}$, the whole orientation state space is accessible,
the cubic cell with $\eta_{cell} =1$ is the only minimiser of $F$, and the
orientational order parameter $S_{2}=0$, the system is isotropic. At $\phi
=\phi _{cp}$, the isotropic state loses its stability and two
orientationally ordered solutions bifurcate from the isotropic branch. The
prolate branch with $S_2>1, \eta_{cell}>1$ has lower energy and pressure
than those of the oblate branch with $S_2<0, \eta_{cell}>1$ for $%
\phi>\phi_{cp}$.

For the oblate ellipsoid with $\eta _{particle}=0.5$, although the
bifurcation is at $\phi _{c}$, orientational order can be observed both
below and above $\phi _{c}$. In fact, there are two first order transitions
from isotropic to both prolate nematic phase and oblate nematic phase at two
different volume fractions slightly below $\phi _{c}$. The prolate branch
with $S_{2}>0,\eta _{cell}<1$ possesses lower energy and pressure than those
of the oblate branch with $S_{2}>0,\eta _{cell}>1$.


For sufficiently large volume fractions, certain orientations are disallowed
by the shape of the cell, and the orientational distribution function has
nontrivial compact support. This is indicated in Fig.~\ref{fig:pdf}(b-e),
where representative equilibrium orientational probability density functions
for a prolate ellipsoid with $\eta _{particle}=2$ in equilibrium prolate $%
\eta _{cell}>1$ (Figs.~\ref{fig:pdf}(b) and (c)) and oblate $\eta _{cell}<1$
(Figs.~\ref{fig:pdf}(d) and (e)) cells are shown. One can readily see that
the distribution function lacks axial symmetry, even in the very dilute
limit where $S_{2}=0$, as shown in Fig.~\ref{fig:pdf}(a). This is because an
uniaxial ellipsoid has more free volume when it aligns along the diagonals
of a rectangular cell compared to other orientations. 

As the volume fraction
is increased, a topological transition takes place: the connected accessible
orientational space becomes disconnected forming a number of disconnected regions, where the orientation of
the ellipsoid is locked in a small region, as can be observed in going from
Figs.~\ref{fig:pdf}(b) to (c) and from (d) to (e). These topological changes also manifest themselves as kinks on the bifurcation diagrams as in Fig.~\ref{fig:bif} and Fig.~\ref{fig:MD}, although the topological change on the prolate branch is not as apparent as the one on the oblate branch.

Since the orientational distribution function is known, in addition to $%
S_{2} $, higher moments $S_{n}=\langle \mathcal{P}_{n}(\mathbf{\hat{m}\cdot 
\hat{e}})\rangle $ can be readily calculated. Here $\mathcal{P}_n$ is the
Legendre polynomial of degree $n$. Although in the isotropic phase $S_{2}=0$%
, interestingly, higher moments may exist as nonzero functions of the
occupied volume fraction $\phi $, as shown in Fig.~\ref{fig:MD}(b). Given the edges and corners of the box, it
is not surprising that the distribution function has higher moments which
are nonzero at all volume fractions. 

The free volume of an arbitrarily shaped particle in a rectangular cell can
be calculated using a support function based approach, where the support
function plays the role of the function $h$ in Eq.~\eqref{eq_V_free}. We
also remark that as long as the cell is convex, the free volume for a
concave particle will be identical to that of its convex hull. Applying the
support function approach to extend our model to particles with more complex
shapes has already been initiated by one of us (JMT).

We realize that cell shapes other than rectangular boxes may allow more free
volume to ellipsoidal particles; we expect that to every particle shape
there corresponds an optimal cavity shape with the lowest free energy in
equilibrium. The rectangular box is just our first attempt; an ellipsoidal
cell is considered in Section \ref{Sec_Needle}.

\subsection{Molecular dynamics}

Molecular dynamics (MD) is a useful tool for studying properties of systems
of interacting particles. MD simulations keep track of the motion of
individual particles relying on classical mechanics. The results of MD
simulations can be used to test the validity of statistical mechanical
models. Here, we perform MD simulations of a hard rigid ellipsoidal particle
moving in a rectangular cell, undergoing collisions with the cell walls. We
imagine that this approximates the real physical situation where thermally
excited molecules collide with nearest neighbors. As we show in this
subsection, MD gives the same predictions for the equilibrium shape of the
cell, for the average orientation of the ellipsoid, and for the pressure as
our statistical mechanical model.

The dynamics consists of free flights and collisions with cell walls. When
the particle does not undergo a collision, the position of its center of
mass is updated as 
\begin{equation}
\mathbf{r(}t_{i+1})=\mathbf{r}(t_{i})+\mathbf{v}_{cm}\Delta t_{i}.
\end{equation}%
However, the angular velocity of the particle changes in time even without
external torque. The angular acceleration in an inertial reference frame is
given by Euler's equations for rigid body dynamics,%
\begin{equation}
\dot{\boldsymbol{\omega }}(t_{i})=\mathbf{-I}^{-1}\cdot (\boldsymbol{\omega }%
\times (\mathbf{I}\cdot \boldsymbol{\omega })),
\end{equation}%
where $\mathbf{I}$ is the tensor of inertia of the ellipsoid, given by 
\begin{equation}
\mathbf{I}=(\frac{2}{5}mb^{2}-\frac{1}{5}m(a^{2}+b^{2}))\mathbf{\hat{m}\hat{m%
}}+\frac{1}{5}m(a^{2}+b^{2})\mathbb{I}\mathbf{,}
\end{equation}%
where $m$ is its uniformly distributed mass. We then update the angular
velocity and the orientation of the particle as follows, 
\begin{eqnarray}
\boldsymbol{\omega }(t_{i}) &=&\boldsymbol{\omega }(t_{i-1})+\boldsymbol{%
\dot{\omega}}\Delta t_{i}, \\
\mathbf{\hat{m}}(t_{i+1}) &=&\frac{\mathbf{\hat{m}}(t_{i})+\boldsymbol{%
\omega }(t_{i})\times \mathbf{\hat{m}}(t_{i})\Delta t_{i}}{|\mathbf{\hat{m}}%
(t_{i})+\boldsymbol{\ \omega }(t_{i})\times \mathbf{\hat{m}}(t_{i})\Delta
t_{i}|},
\end{eqnarray}%
where we use the default value of $\Delta t_{i}=10^{-3}$. If the updated
configuration causes any part of the particle to leave the cell, we return
to the previous step, and halve $\Delta t_{i}$. The process is repeated
until the new configuration of the particle is completely within the cell.
When the time step size reaches a lower threshold (e.g., $\Delta
t_{i}=10^{-14}$), collision with the wall occurs.

We assume that both the particle and cell walls are frictionless, thus the
impulses are along the normals to the surfaces in contact. In a collision,
the conservation of linear momentum, angular momentum and kinetic energy,
gives the postcollision center of mass velocity $\mathbf{v}_{f}$ and angular
velocity $\boldsymbol{\omega }_{f}$ by (\cite{paradox})%
\begin{equation}
\mathbf{v}_{f}=\mathbf{v}_{i}-\frac{J}{m}\mathbf{\hat{N}},
\end{equation}%
\begin{equation}
\boldsymbol{\omega }_{f}=\boldsymbol{\omega }_{i}-J\mathbf{I}^{-1}\cdot (%
\mathbf{p\times \hat{N}}),
\end{equation}%
where $\mathbf{v}_{i}$ and $\boldsymbol{\omega }_{i}$ are precollision
velocity and angular velocity, $\mathbf{\hat{N}}$ is the outward normal of
the ellipsoid pointing towards the cell wall. The vector $\mathbf{p}$ is a
body-fixed vector joining the center of mass of the ellipsoid to the point
of contact with the cell wall at the instant of collision, 
\begin{equation}
\mathbf{p}=\frac{\mathbf{A}^{-1}\cdot \mathbf{\hat{N}}}{\sqrt{\mathbf{\hat{N}%
}\cdot \mathbf{A}^{-1}\cdot \mathbf{\hat{N}}}}
\end{equation}%
and the positive scalar $J$ is the magnitude of the impulse transferred to
the ellipsoid at the collision, 
\begin{equation}
J=2\frac{(\mathbf{\mathbf{v}}_{i}+\mathbf{\mathbf{\omega }}_{i}\times 
\mathbf{p)\cdot \hat{N}}}{1/m-\mathbf{\hat{N}}\cdot \mathbf{p}\times \mathbf{%
I}^{-1}\times \mathbf{p}\cdot \mathbf{\hat{N}}}.  \label{6}
\end{equation}

The pressure on each wall is the average impulse per area on the wall. If
the pressure happens not to be isotropic, we then update the relevant length
of the cell, denoted $d_l$, according to 
\begin{equation}
d_{l}\rightarrow d_{l}+\gamma \frac{P_{l}-P_{t}}{P_{l}},
\end{equation}%
where $P_{l}$ and $P_{t}$ represent the averaged pressures on two square
walls and four rectangular walls, respectively, and $\gamma =0.01$ plays the
role of compressibility. The updates for the other two dimensions are
determined by maintaining constant volume and a uniaxial shape.

A very great deal of work has been done on simulations of systems of hard
ellipsoids\cite{Frenkel},\cite{Perram},\cite{odriozola},\cite{Berne}. Our
system is different, however, due to the mean field aspect of our model,
where only a single particle is present, enclosed by a cell representing the
effects of the other particles. Confinement induced ordering of hard
ellipsoids has also been studied between two hard walls \cite{Mao} 
however our confinement is in 3D; furthermore the shape of the confining shape
is adaptive.

The scalar order parameter is calculated according to $S_{n}=\langle \mathcal{P}_{n}(\mathbf{\hat{m}\cdot 
	\hat{e}})\rangle $
where the average is over all instantaneous orientations of the symmetry
axis $\mathbf{\hat{m}}$ of the ellipsoid between collisions. All simulation
results were averaged over $10^{3}$ random initial conditions.

\begin{figure}[h]
	\centering
\begin{subfigure}[c]{0.4\linewidth}
	\centering
	\includegraphics[width=\linewidth]{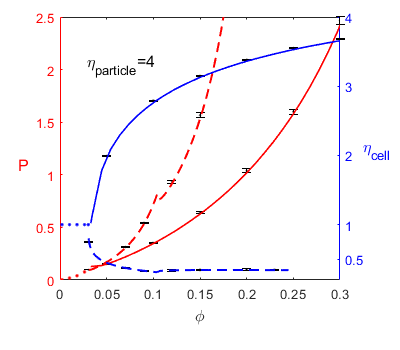}\caption{}
\end{subfigure}
\qquad
\begin{subfigure}[c]{0.4\linewidth}
	\centering
	\includegraphics[width=\linewidth]{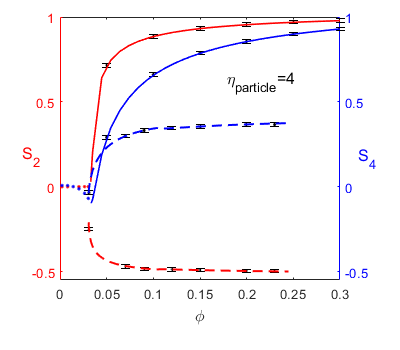}\caption{}
\end{subfigure}
\caption{ Results for ellipsoids with aspect ratio $\protect\eta_{particle}
=4$. (a) Pressure $P$ (left axes) and cell aspect ratio $\protect\eta_{cell}$ (right axis) vs.~volume fraction $\protect\phi$. (b) Scalar order parameters $S_2$ (left axes) and $S_4$ (right axes) vs.~volume fraction $\phi$. Only locally stable phases are shown. Solid
lines are prolate phase, and dashed lines are for oblate phase, and dotted lines for isotropic phase.  The MD results are shown as discrete points
with error bars.}
\label{fig:MD}
\end{figure}

Figure \ref{fig:MD} shows the agreement of the results of free
volume calculation in Sect.~\ref{Sec_free_vol} and MD simulations for an
ellipsoid with aspect ratio $4$. The transition from the isotropic phase to the prolate phase is continuous, whereas the transition from the isotropic phase to the oblate phase is discontinuous. This is different from the case of $\eta_{particle}=2$ where both transitions are continuous. 


\subsection{Needle in an ellipsoidal cell}

\label{Sec_Needle}

In this subsection, we consider a needle within an ellipsoidal cell. This
geometry allows for an analytic free volume expression, and reveals a
topology common to the case of very slender particles.

\begin{figure}[h]
\centering
\begin{subfigure}[c]{0.3\linewidth}
		\centering
		\includegraphics[width=0.7\linewidth]{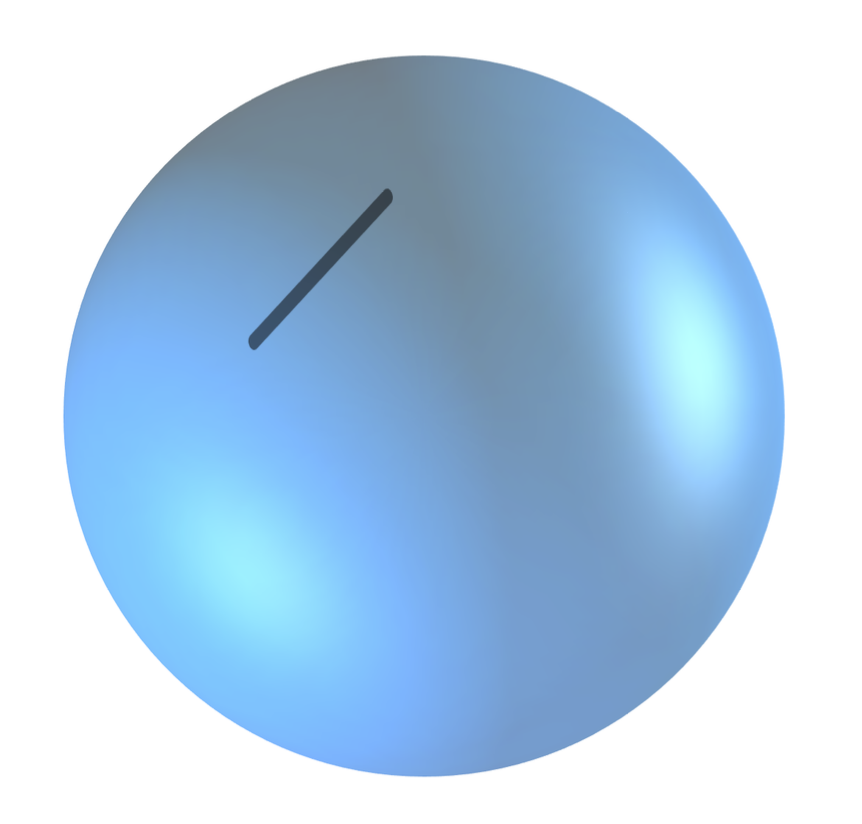}
		\caption{}
		\label{fig:sphere}
	\end{subfigure}
\quad\quad 
\begin{subfigure}[c]{0.3\linewidth}
		\centering
		\includegraphics[width=.4\linewidth]{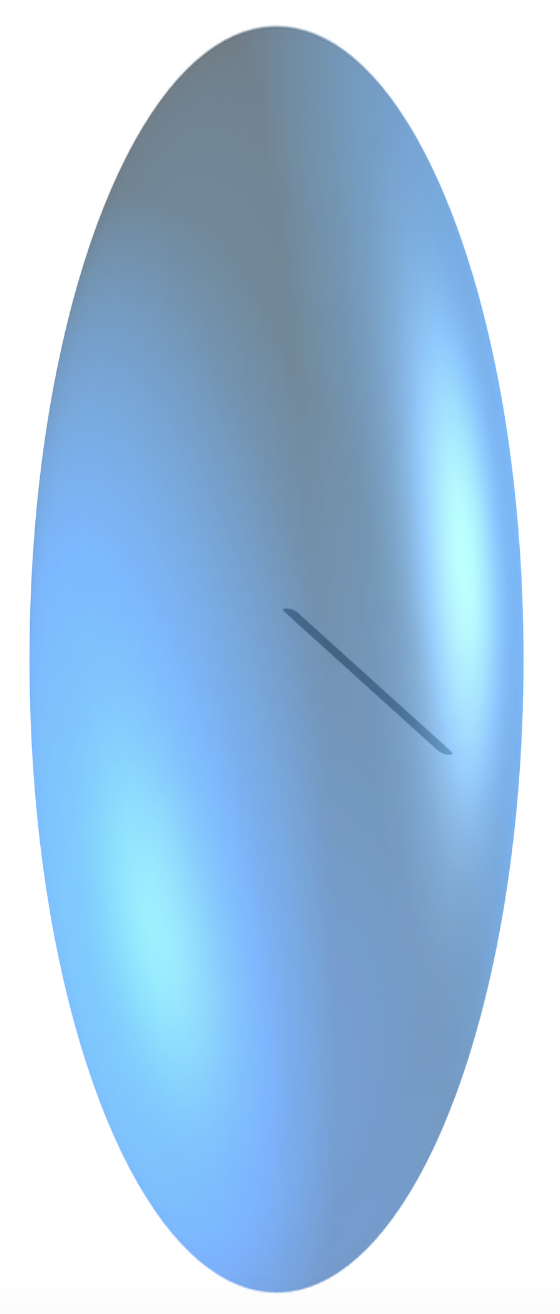}
		\caption{}
		\label{fig:ellipsoid}
	\end{subfigure}
\caption{The undeformed spherical globule $\mathcal{C}_0$ enclosing a rigid
rod (a) and its ellipsoidal image $\mathcal{C}$ under affine deformation
(b). }
\end{figure}

Consider $\mathcal{C}_0$ to be a spherical cavity $\mathscr{B}_R$ of radius $%
R$ enclosing a rigid rod of length $2a$ (see Fig.~\ref{fig:sphere}) and
subject the sphere to a homogeneous, isochoric deformation $\boldsymbol{f}$
with gradient $\mathbf{F}$, which, modulo a rotation, changes the sphere
into the ellipsoid $\mathcal{C}$ depicted in Fig.~\ref{fig:ellipsoid}.

Now, the unit vector $\mathbf{\hat{m}}$ denotes the orientation of the rod,
such that if the center of the rod is at the point $\boldsymbol{y}$, its
ends occupy the points $\boldsymbol{y}_{\pm }=\boldsymbol{y}\pm a\mathbf{%
\hat{m}}$, respectively. We want to determine the volume of the free region $%
\mathscr{R}_{\mathrm{f}}(\mathbf{\hat{m}})$ accessible to $\boldsymbol{y}$
within the ellipsoid $\mathcal{C}$, for a given orientation $\mathbf{\hat{m}}
$ of the rod. Writing $\boldsymbol{y}_{\pm }=\mathbf{F}\boldsymbol{x}_{\pm }$
and $\boldsymbol{y}=\mathbf{F}\boldsymbol{x}$, so that $\boldsymbol{x}_{\pm
} $ are the preimages of $\boldsymbol{y}_{\pm }$, and $\boldsymbol{x}$ the
preimage of $\boldsymbol{y}$, the region $\boldsymbol{f}^{-1}(\mathscr{R}_{%
\mathrm{f}})$ comprises all points $\boldsymbol{x}$ such that 
\begin{equation}
\mathbf{F}\boldsymbol{x}\pm a\mathbf{\hat{m}}=\mathbf{F}\boldsymbol{x}_{\pm
}\quad \text{for both}\quad \boldsymbol{x}_{\pm }\in \mathscr{B}_{R},
\label{eq:images}
\end{equation}%
from which, since $\mathbf{F}$ is invertible, we have that 
\begin{equation}
\boldsymbol{x}\pm a\mathbf{F}^{-1}\mathbf{\hat{m}}=\boldsymbol{x}_{\pm
},\quad \boldsymbol{x}_{\pm }\in \mathscr{B}_{R},  \label{eq:preimages}
\end{equation}%
and so 
\begin{equation}
\boldsymbol{f}^{-1}(\mathscr{R}_{\mathrm{f}})=\{\boldsymbol{x}:|\boldsymbol{x%
}+a\mathbf{F}^{-1}\mathbf{\hat{m}}|\leq R\}\cap \{\boldsymbol{x}:|%
\boldsymbol{x}-a\mathbf{F}^{-1}\mathbf{\hat{m}}|\leq R\}.
\label{eq:preregion}
\end{equation}%
Geometrically, this set is the region comprised of two spheres of radius $R$
whose centers are $2b$ apart, where 
\begin{equation}
b:=a|\mathbf{F}^{-1}\mathbf{\hat{m}}|.  \label{eq:b_definition}
\end{equation}%
Since $\det \mathbf{F}=1$, the free volume $V_{\mathrm{f}}$ is the same as
the volume of $\boldsymbol{f}^{-1}(\mathscr{R}_{\mathrm{f}})$, that is,
twice the volume of a spherical frustum of radius $r:=\sqrt{R^{2}-b^{2}}$
and height $h:=R-b$, 
\begin{equation}
V_{\mathrm{f}}=\frac{\pi h}{3}(3r^{2}+h^{2})=V_{cell}\left( 1-\frac{3}{2}%
\alpha \beta +\frac{1}{2}\alpha ^{3}\beta ^{3}\right) ,
\label{eq:free_volume}
\end{equation}%
where $V_{cell}$ is the volume of $\mathscr{B}_{R}$ and we have set 
\begin{equation}
\alpha =\frac{a}{R}\leq 1\quad \text{and}\quad \frac{b}{R}=\alpha \beta \leq
1.  \label{eq:alpha_beta}
\end{equation}%
It readily follows from \eqref{eq:preregion} and \eqref{eq:free_volume} that 
\begin{equation}
\beta =\sqrt{\mathbf{\hat{m}}\cdot \mathbf{B}^{-1}\mathbf{\hat{m}}},
\label{eq:beta}
\end{equation}%
where $\mathbf{B}=\mathbf{F}\mathbf{F}^{\mathsf{T}}$ is the left
Cauchy-Green tensor, which in the frame $(\mathbf{\hat{e}}_{1},\mathbf{\hat{e%
}}_{2},\mathbf{\hat{e}}_{3})$ of the principal directions of stretching is
represented as 
\begin{equation}
\mathbf{B}=\lambda _{1}^{2}\mathbf{\hat{e}}_{1}\mathbf{\hat{e}}_{1}+\lambda
_{2}^{2}\mathbf{\hat{e}}_{2}\mathbf{\hat{e}}_{2}+\lambda _{3}^{2}\mathbf{%
\hat{e}}_{3}\mathbf{\hat{e}}_{3},  \label{eq:B_representation}
\end{equation}%
where $\lambda _{1}$, $\lambda _{2}$, and $\lambda _{3}$ are the principal
stretches subject to $\lambda _{1}\lambda _{2}\lambda _{3}=1$. Letting $%
m_{i} $ be the components of $\mathbf{\hat{m}}$ in the frame $(\mathbf{\hat{e%
}}_{1},\mathbf{\hat{e}}_{2},\mathbf{\hat{e}}_{3})$, they can be written as 
\begin{equation}
m_{1}=\sin \theta \cos \varphi ,\quad m_{2}=\sin \theta \sin \varphi ,\quad
m_{3}=\cos \theta ,  \label{eq:n_i_components}
\end{equation}%
for $\theta \in \lbrack 0,\pi ]$ and $\varphi \in \lbrack 0,2\pi ]$. By use
of \eqref{eq:n_i_components} and \eqref{eq:B_representation} in %
\eqref{eq:beta} , we arrive at 
\begin{equation}
\beta =\sqrt{\frac{1}{\lambda _{1}^{2}}\sin ^{2}\theta \cos ^{2}\varphi +%
\frac{1}{\lambda _{2}^{2}}\sin ^{2}\theta \sin ^{2}\varphi +\frac{1}{\lambda
_{3}^{2}}\cos ^{2}\theta }.  \label{eq:beta_general}
\end{equation}%
For simplicity, we shall assume that $\mathbf{B}$ has the uniaxial symmetry
around $\hat{\mathbf{e}}_{3}$, so that $\lambda _{1}=\lambda _{2}=1/\sqrt{%
\lambda _{3}}$ and $\beta $ reduces to 
\begin{equation}
\beta =\sqrt{{\lambda }\sin ^{2}\theta +\frac{1}{\lambda ^{2}}\cos
^{2}\theta },\text{ with }\lambda =\lambda _{3}.  \label{eq:beta_reduced}
\end{equation}%
With this choice, the ellipsoid is prolate along $\hat{\mathbf{e}}_{3}$ if $%
\lambda >1$ and oblate if $\lambda <1$.

For given $\lambda$, the domain $I_{\lambda}$ of admissible values of $%
\theta $ is restricted by the second inequality in \eqref{eq:alpha_beta}. A
direct inspection shows that where 
\begin{equation}
I_{\lambda }=\left\{ 
\begin{array}{cc}
\lbrack \chi ,\pi -\chi ], & 0\leq \lambda\leq \alpha \\ 
\lbrack 0,\pi ], & \alpha \leq \lambda \leq \frac{1}{\alpha^2 } \\ 
\lbrack 0,\chi ]\cup \lbrack \pi -\chi ,\pi ], & \lambda\geq \frac{1}{%
\alpha^2 }%
\end{array}%
\right.
\end{equation}%
where%
\begin{equation}
\chi =\arccos \left(\frac{\lambda}{\alpha}\sqrt{\frac{1-\lambda \alpha^2 }{%
1-\lambda^3}}\right).  \label{eq:chi}
\end{equation}

\begin{figure}[h]
\centering
\begin{subfigure}[c]{0.3\linewidth}
		\centering
		\includegraphics[width=\linewidth]{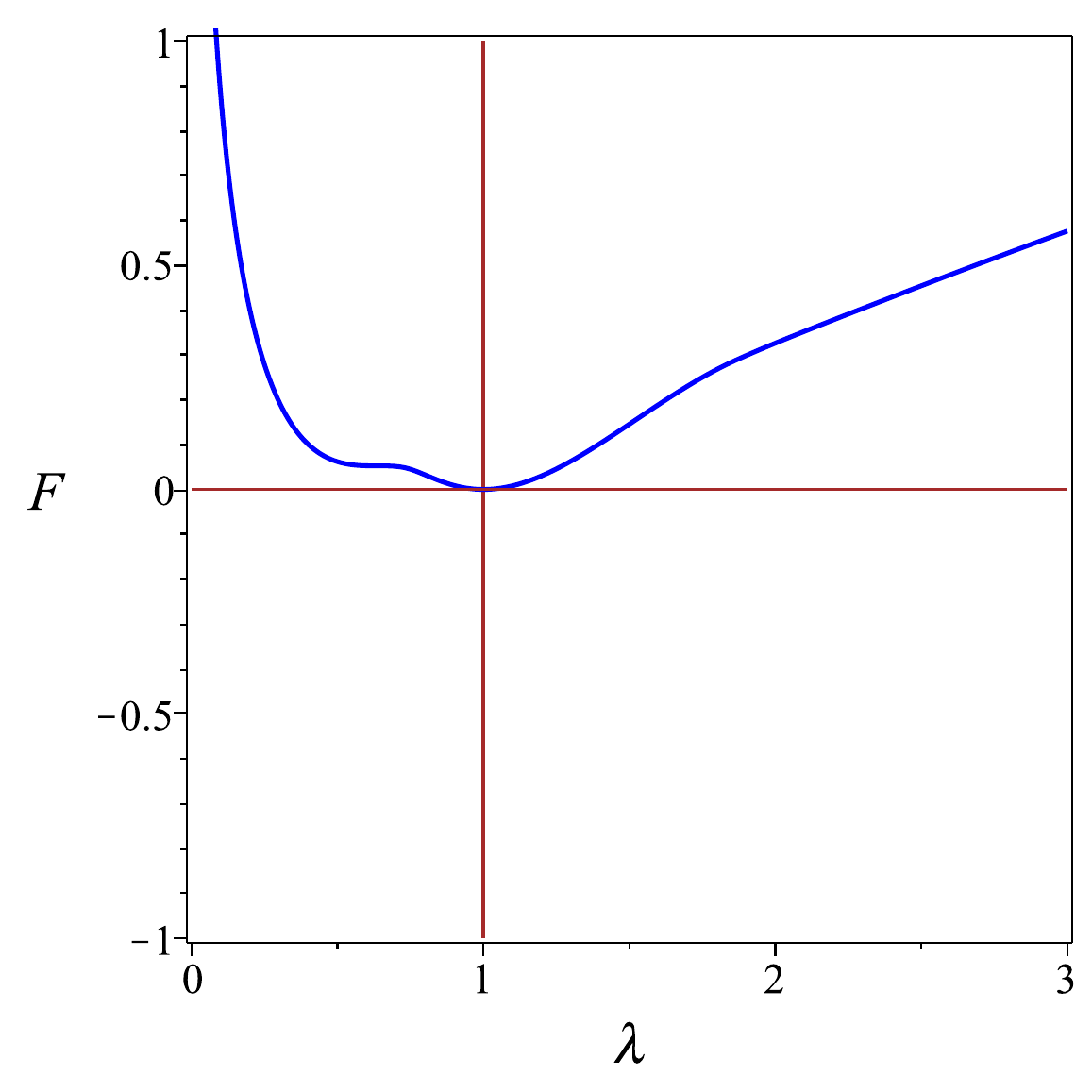}
		\caption{$\alpha=\alpha_c^{(1)}\dot{=}0.737$}
		\label{fig:f_1}
	\end{subfigure}
\quad 
\begin{subfigure}[c]{0.3\linewidth}
		\centering
		\includegraphics[width=\linewidth]{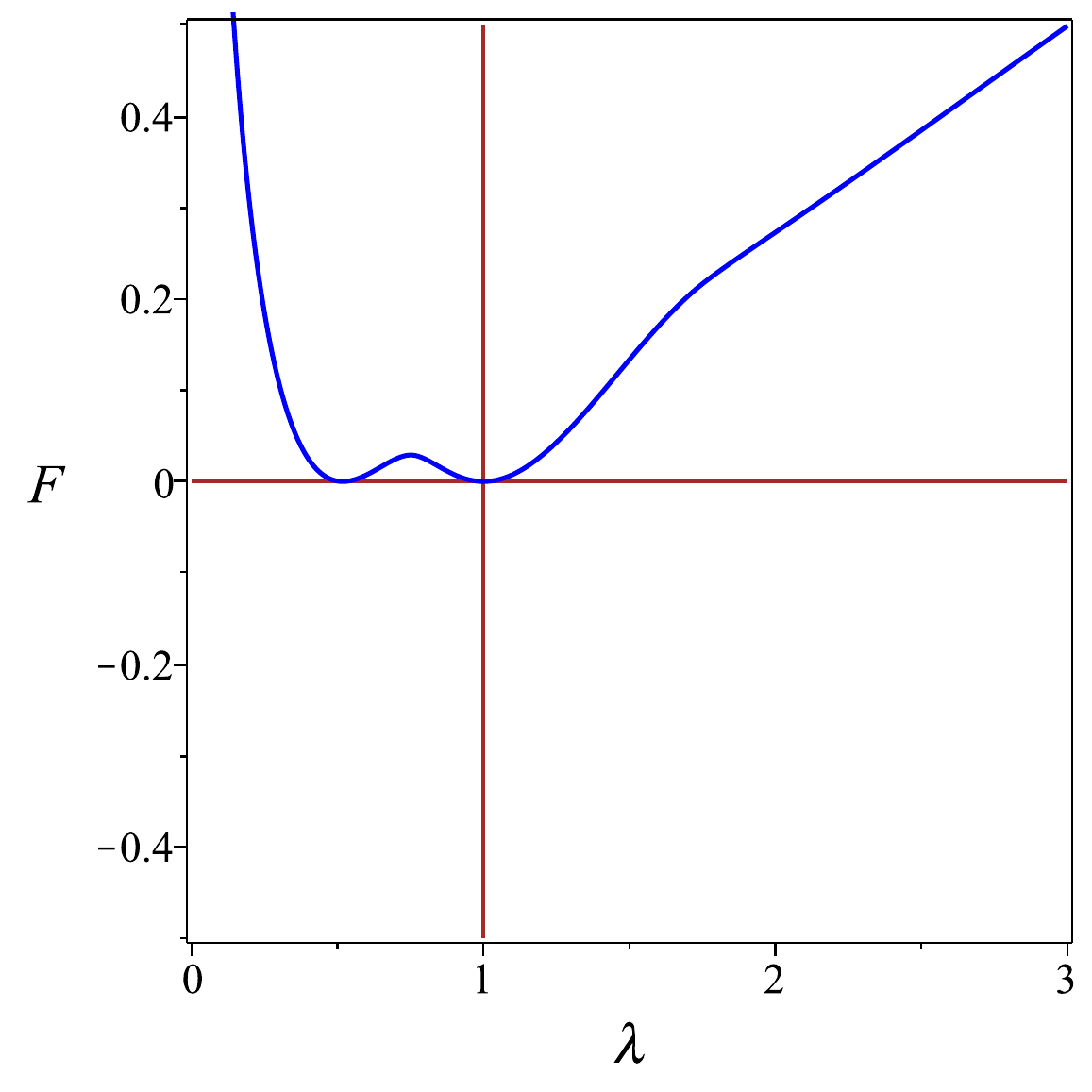}
		\caption{$\alpha=\alpha_t\dot{=}0.757$}
		\label{fig:f_2}
	\end{subfigure}
\quad 
\begin{subfigure}[c]{0.3\linewidth}
		\centering
		\includegraphics[width=\linewidth]{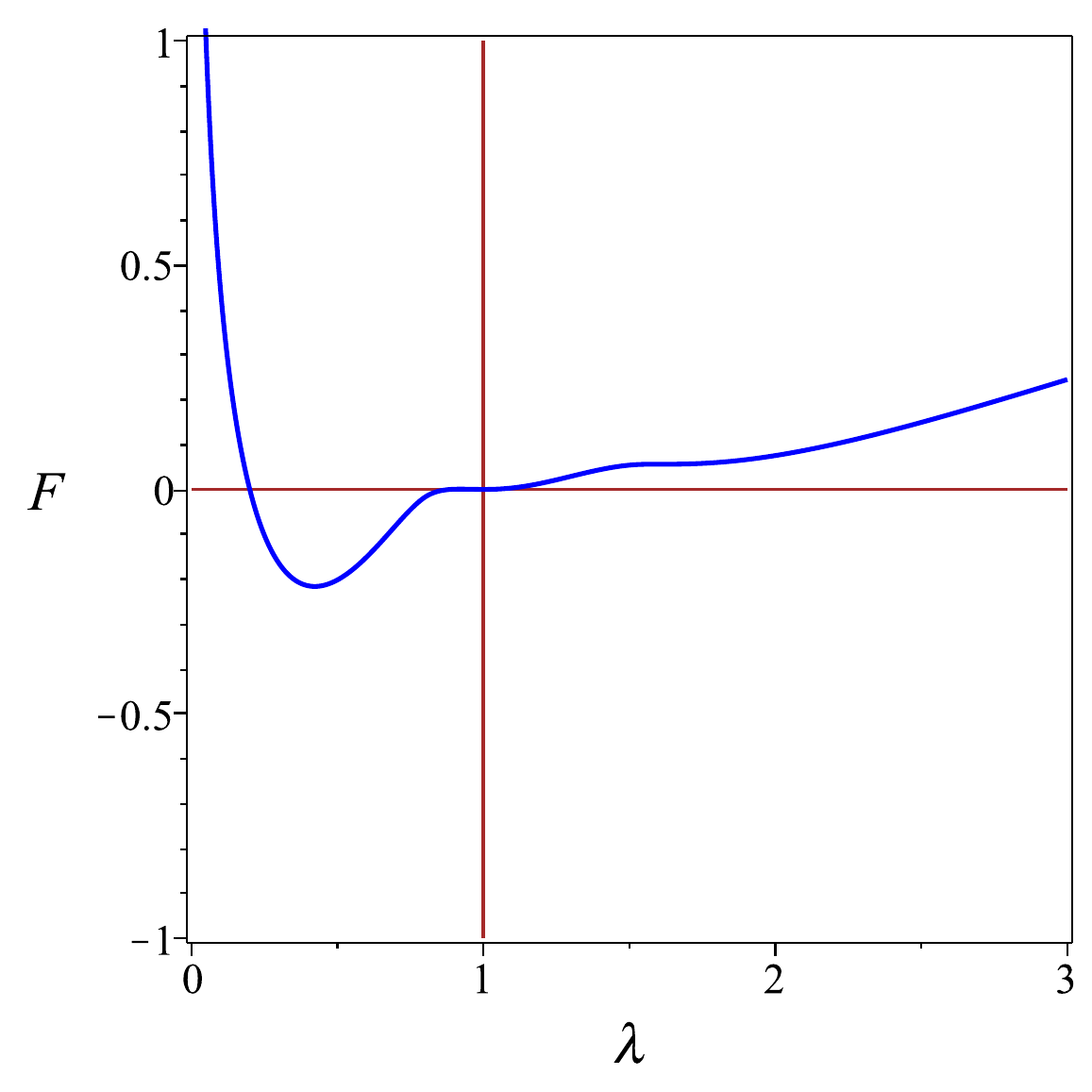}
		\caption{$\alpha=\alpha_c^{(2)}\dot{=}0.805$}
		\label{fig:f_3}
	\end{subfigure}
\newline
\begin{subfigure}[c]{0.3\linewidth}
		\centering
		\includegraphics[width=\linewidth]{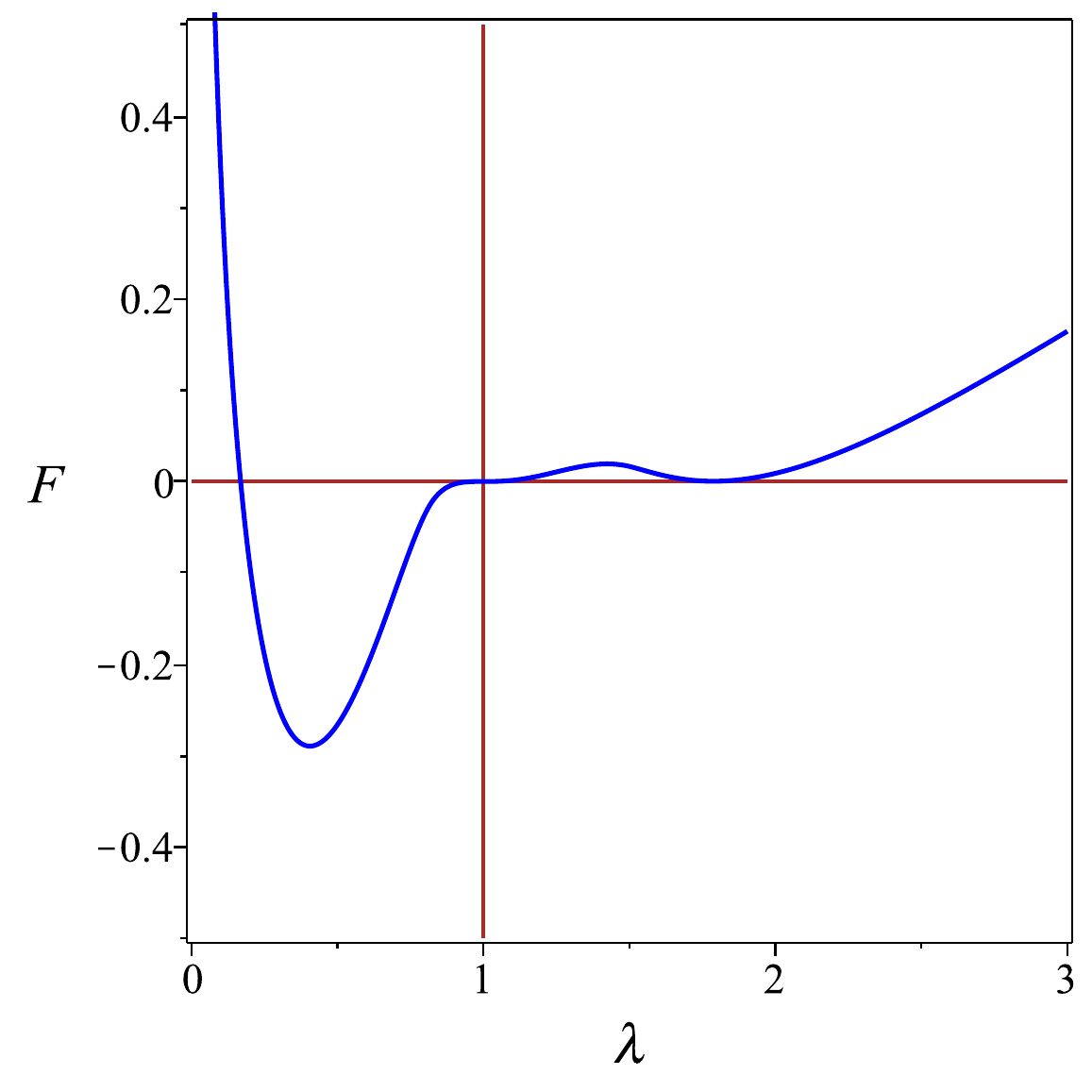}
		\caption{$\alpha=\sqrt{2/3}\dot{=}0.816$}
		\label{fig:f_4}
	\end{subfigure}
\quad 
\begin{subfigure}[c]{0.3\linewidth}
		\centering
		\includegraphics[width=\linewidth]{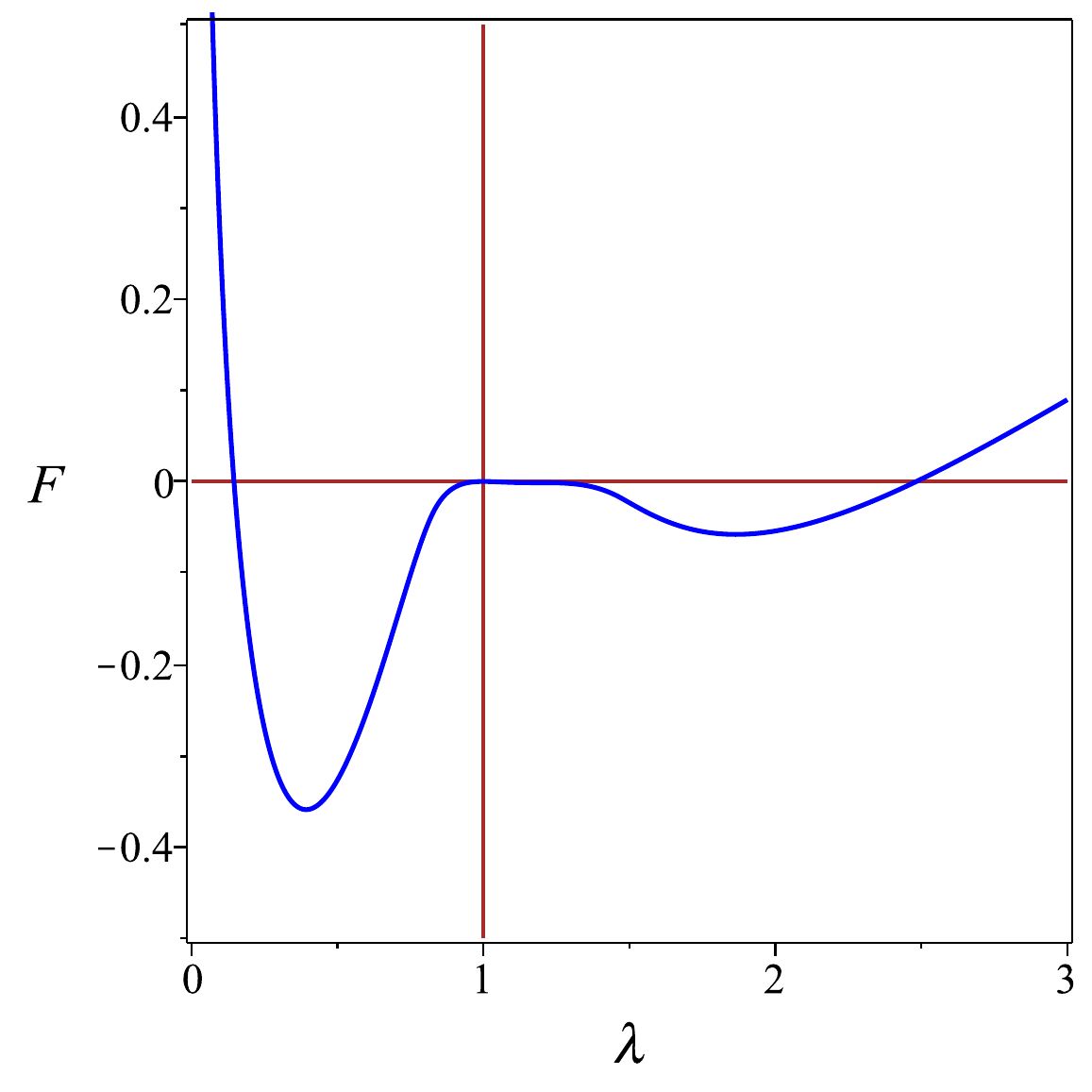}
		\caption{$\alpha=\alpha_c^{(3)}\dot{=}0.826$}
		\label{fig:f_5}
	\end{subfigure}
\quad 
\begin{subfigure}[c]{0.3\linewidth}
		\centering
		\includegraphics[width=\linewidth]{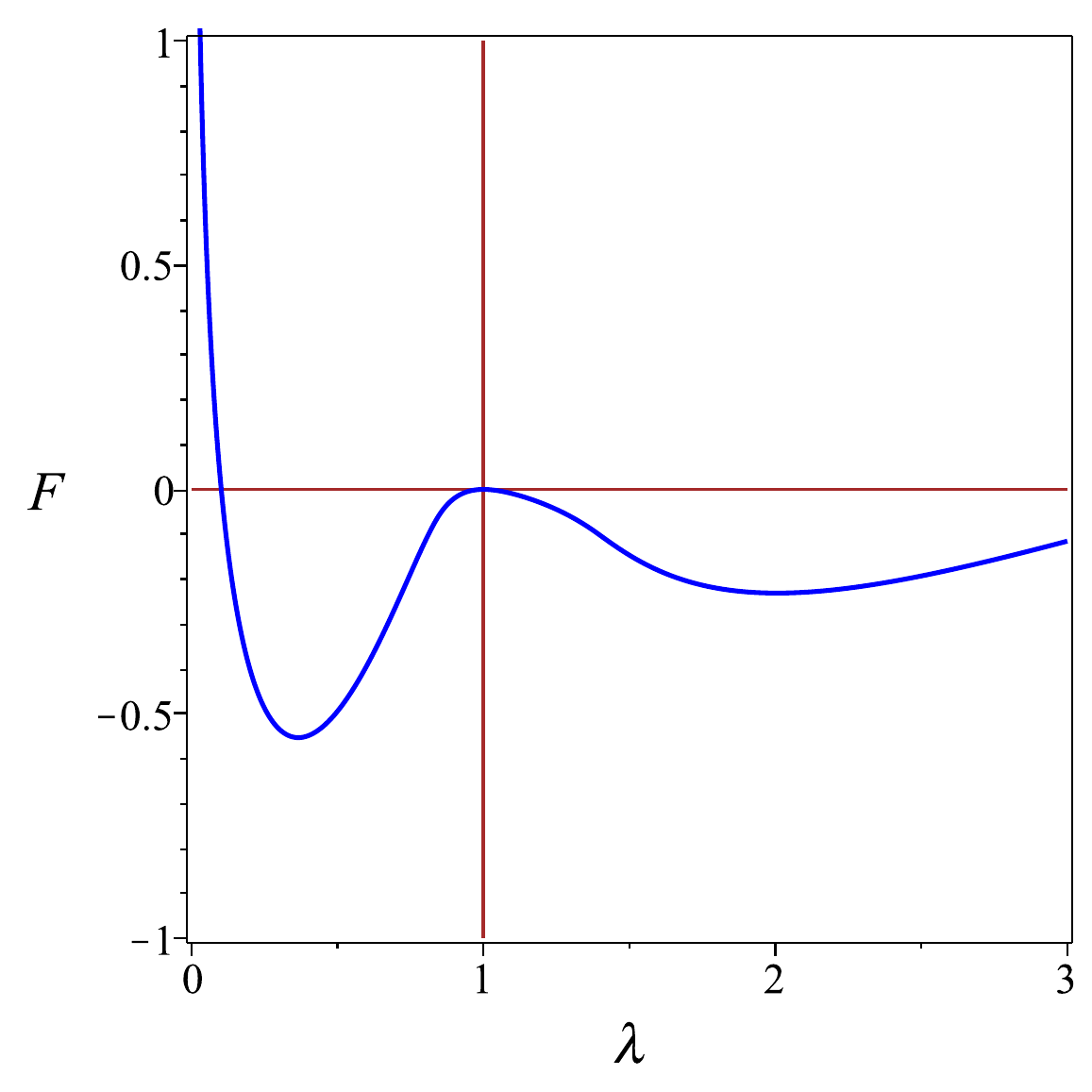}
		\caption{$\alpha=0.850$}
		\label{fig:f_6}
	\end{subfigure}
\caption{The dimensionless free energy $f$ as in \eqref{eq:f} plotted
against $\protect\lambda $ for different values of $\protect\alpha $.}
\label{fig:f_s}
\end{figure}

The free energy of the rod (in units $kT$) is then given by 
\begin{equation}
F(\alpha ,\lambda)=-\ln \left( \pi \int_{I_{\lambda}}(2-3\alpha \beta
+\alpha ^{3}\beta ^{3})\sin \theta d\theta \right) +\ln [2\pi (\alpha
-1)^{2}(\alpha +2)],  \label{eq:f}
\end{equation}%
where the added constant has been gauged so as to ensure that $F(\alpha
,1)=0 $. The integral in \eqref{eq:f} can be expressed in terms of
elementary functions, not given explicitly here in the interest of space.
Plots of $F$ as function of $\lambda$ are shown in Fig.~\ref{fig:f_s} for
several values of $\alpha $. 

For $\alpha <\alpha _{\mathrm{c}}^{(1)}\doteq 0.737$, $F$ has a single
critical point at $\lambda =1$, its absolute minimum representing the
isotropic state where the spherical globule remains undeformed. For $\alpha
>\alpha _{\mathrm{c}}^{(1)}$, a local minimum and a local maximum develop in
the oblate branch (where $\lambda <1$), while the isotropic state remains
the absolute minimum, until $\alpha $ reaches the transition value $\alpha _{%
\mathrm{t}}\doteq 0.757$, where both the isotropic state and the oblate
minimum have the same energy. A first-order transition takes place at $%
\alpha =\alpha _{\mathrm{t}}$, with the oblate minimum at $\lambda \doteq
0.518$. At $\alpha =\alpha _{c}^{(2)}\dot{=}0.805$, an inflection point
develops on the prolate branch, from which a local maximum and a local
minimum emanate, the former approaching the equilibrium isotropic state at $%
\lambda =1$. At $\alpha =\sqrt{\frac{2}{3}}\doteq 0.816$, the isotropic
state becomes unstable and two branches emanate from it, a local prolate
minimum and a local oblate maximum: the former connects with the prolate
maximum at $\alpha =\alpha _{c}^{(3)}\dot{=}0.826$, while the latter
connects with the oblate minimum at $\alpha =\alpha _{c}^{(1)}$. The locally
stable prolate branch converges to $\lambda \doteq 2.627$ as $\alpha
\rightarrow 1$, whereas the globally stable oblate branch converges to $%
\lambda \dot{=} 0.254$ as $\alpha\to 1$.

\begin{figure}[h]
\centering
\begin{subfigure}[c]{0.4\linewidth}
\includegraphics[width=\linewidth]{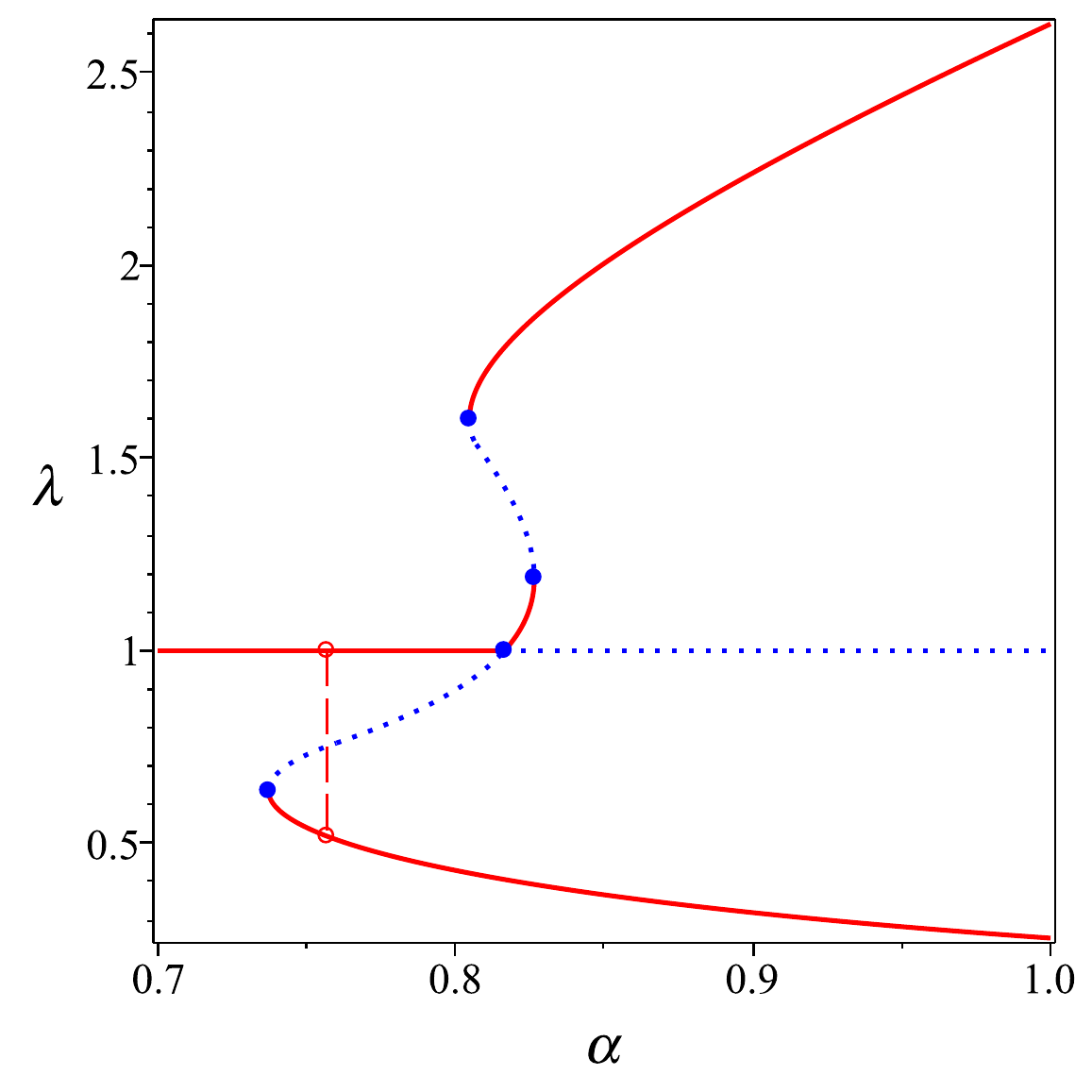}
\caption{}
\end{subfigure} 
\qquad
\begin{subfigure}[c]{0.4\linewidth}
\includegraphics[width=\linewidth]{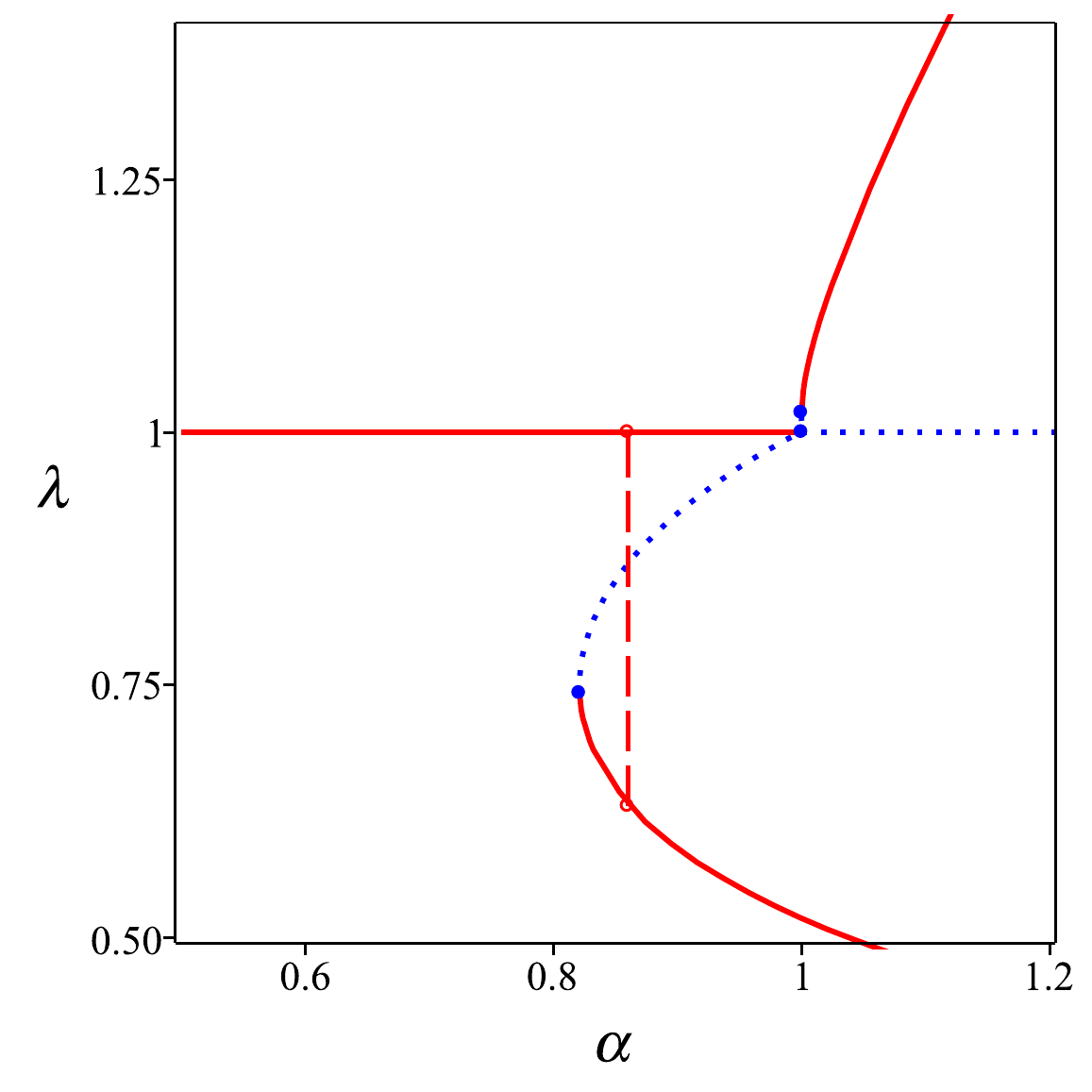}\caption{}
\end{subfigure}
\caption{Bifurcation diagrams. (a) Needle in an ellipsoid cell. (b) Needle
in a rectangular cell. In the former case $\protect\alpha = 2a/\sqrt[3]{%
3V_{cell}/(4\protect\pi)}$, while in the latter $\protect\alpha=2a/\sqrt[3]{%
V_{cell}}$. Solid red lines are local minima, dotted blue lines are local
maxima. Red circles mark the transition from the isotropic state to the
absolute oblate minimiser. Blue circles represent the critical points
discussed in the text.}
\label{fig:bifurcation}
\end{figure}

Figure \ref{fig:bifurcation} summarizes the above results, which are also
compared with the bifurcation diagrams for a needle in a rectangular cell.
We note that the topologies are mildly distinct. Whilst the oblate branch has a similar topology in each case, two unstable branches, one prolate and one oblate, bifurcate from the isotropic state at the critical value $\alpha=1$ in the cuboidal case, with the former being a very small region that is difficult to see in the figure. 

\section{Work by shape change}

\label{Sec_Disscussion} Our mean field model gives some insights about the
effects of shape change by individual molecules on their neighbors. A number
of photomechanical materials consist of a nematic liquid crystal elastomer,
containing a small fraction ($\sim 2\%$) of photoisomerizable azo dyes.
These, in their ground state trans- configuration are elongated and in
essence very similar to the liquid crystal molecules of the host. When
illuminated, however, they undergo a trans-cis photoisomerization, and their
shape changes from the elongated trans- form to the more compact -- cis-
form. Based on our model, we can estimate the free energy increase and
stress on the host due to this shape change.

\begin{figure}[h]
\centering
\begin{subfigure}[c]{0.45\linewidth}
		\centering
		\includegraphics[width=\linewidth]{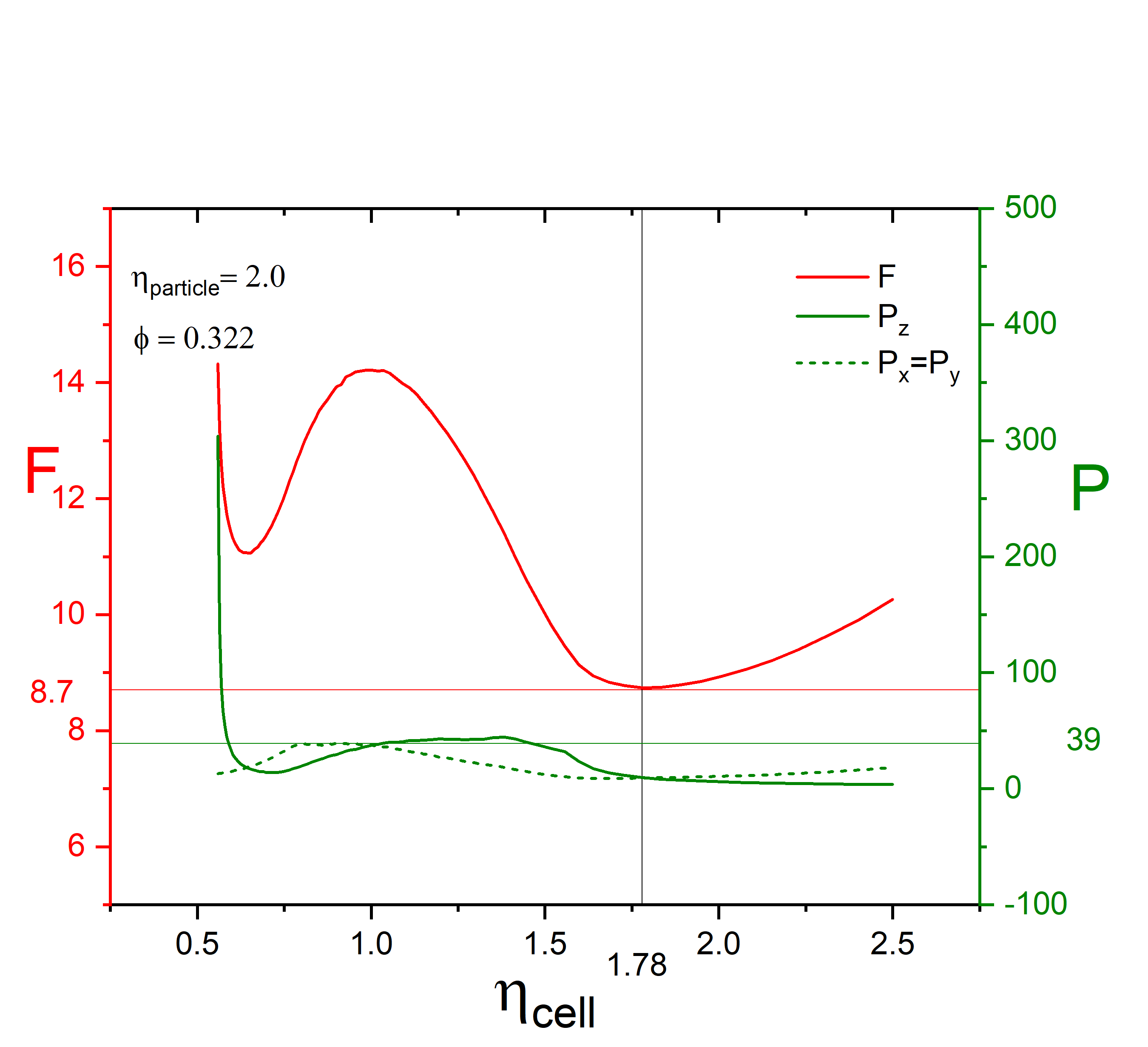}
		\caption{}
		\label{fig:ppm_332_05}
	\end{subfigure}
\quad\quad 
\begin{subfigure}[c]{0.45\linewidth}
		\centering
		\includegraphics[width=\linewidth]{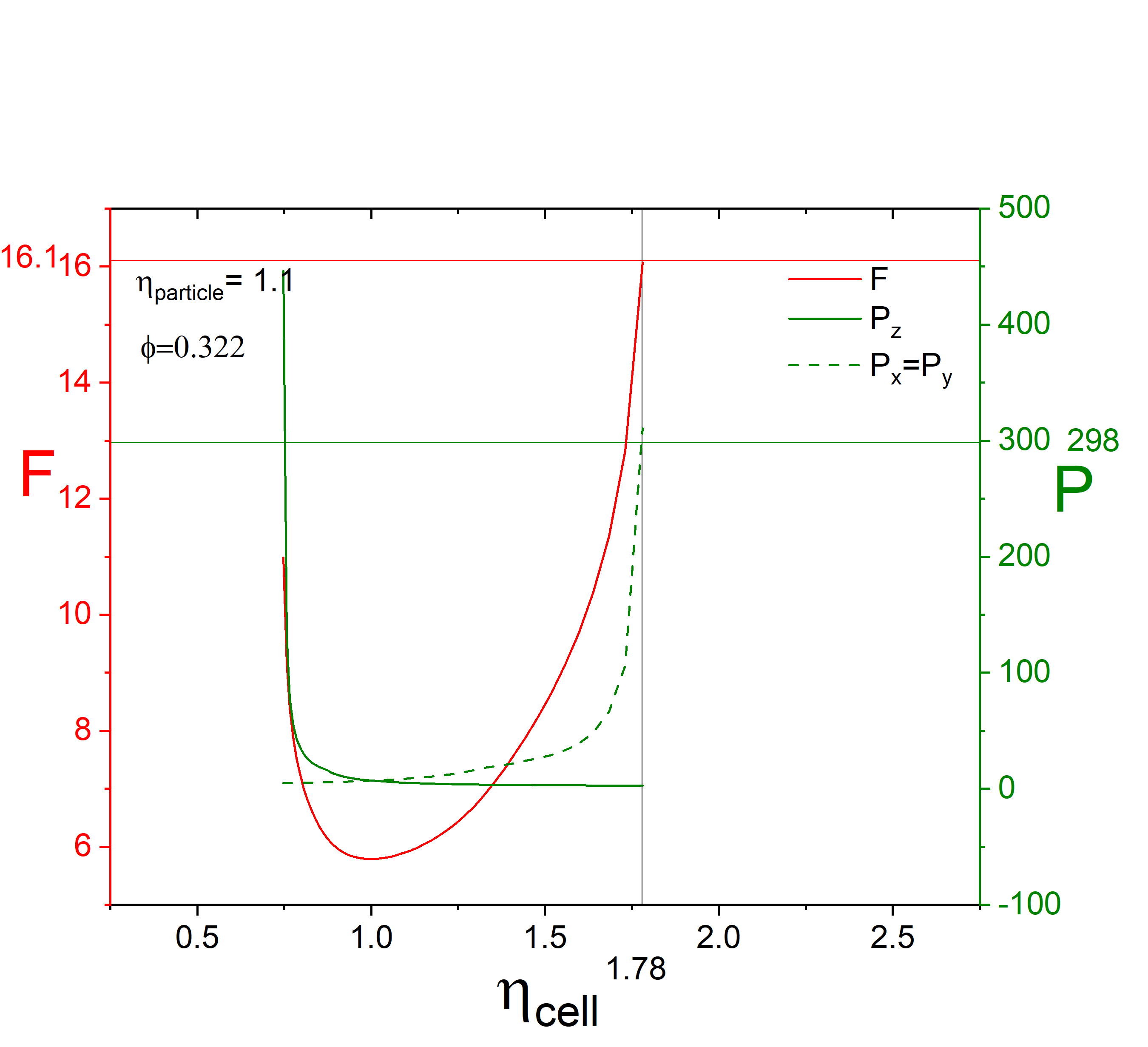}
		\caption{}
		\label{fig:ppm_332_09}
	\end{subfigure}
\caption{The particle shape change (decrease in aspect ratio $\protect\eta_{particle}$%
) at $\protect\phi =0.322$ causes a large ($\sim$ factor of 2) increase in the free energy $F$, and a
large increase ($\sim$ factor of 10) in the pressure perpendicular to the symmetry axis.}
\label{fig:shapes}
\end{figure}

The aspect ratio of the trans- isomer is (roughly) known; we can also
estimate the volume fraction. We know that in equilibrium the free energy is
a minimum; so we minimize the free energy with respect to the cell aspect
ratio. A possible starting configuration is shown in Fig.~\ref{fig:shapes}%
(a); the box shape is optimal for the given volume fraction $\phi =0.322$
and ellipsoid aspect ratio of $2$; the sample is in equilibrium and the
pressure on the box is isotropic. The sample is then illuminated with UV
light, and there is photoisomerization where the dye molecule changes its
shape from the elongated trans- to the more spherical cis- shape. As the
ellipsoid suddenly becomes more spherical, the system loses equilibrium, the
free energy dramatically increases and the pressure becomes anisotropic as
shown in Fig.~\ref{fig:shapes}(b). This is the initial free energy for the
photowork, shown in Fig.~\ref{fig:shapes}(b), together with pressure, which
is now strongly anisotropic.

If allowed, the cell shape then changes to aspect ratio close to $1$ to
minimize the free energy for the new ellipsoid shape. The photomechanical
work is carried out by the stress on the cell walls. The free energy
corresponding to box aspect ratio $1$ is the final energy of the system
after the photowork. The free energy difference between the initial and
final states is the energy available for photowork.

In an isotropic medium, dissolved azo dyes behave like standard two-level
systems, governed by a nearly constant potential energy difference between
the tran- and cis- states. In a liquid crystal host, however, potential
energy of each isomer\footnote{%
internal energy plus the work required to insert the isomer into the
standard cell in the system} would sensitively depend on the anisotropy of
the cell. One would expect therefore that the temperature dependence of
isomer populations in liquid crystals to significantly differ from that of a
simple two-level system.

Energy for the photoexcitation comes from the absorbed photon; what is not
used in increasing the free energy is released as heat. Matching the free
energy increase to the absorbed photon energy may be a viable strategy to
increase the efficiency of photomechanical materials.

\section{Conclusion}

\label{Sec_Conclusion}

In this work, we have considered a mean field model of $N$ indistinguishable
particles with hard core steric interactions in a volume $V$ at temperature $%
T$. The region is divided into $N$ identical cells, each with one particle
in each cell. Each cell has a volume $V/N$. At nonzero temperatures, the
particles undergo collisions with the cell walls which approximate the
effects of neighboring particles. In equilibrium, satisfying
self-consistency, the cell adopts a shape which minimizes the free energy,
and leads to an isotropic pressure tensor.

We have applied this mean field cell model to the case of hard uniaxial
ellipsoids in rectangular and ellipsoidal cells.

At low occupied volume fractions, the shortest dimension of the cell is
longer than the longest dimension of the particles. Hence all orientations
of the particles are possible in the cells; here the particles are
orientationally disordered, and the orientational order parameter - the
second moment of the orientational distribution - is zero. As the occupied
volume fraction increases, there is a critical volume fraction where the
shortest dimension of the cell is equal to the longest dimension of the
particles. Above this critical volume fraction, the orientational
distribution function has compact support; there is a bifurcation and the
system becomes orientationally ordered. One of the orientationally ordered
solutions is prolate, while the other is oblate; the stability is determined
by the shape of the cell. The results of mean-field theory are fully
compatible with the the results of molecular dynamics simulations.

The above results give insights towards understanding the shape change
exhibited by photomechanical materials. In such azo-dye doped liquid crystal
elastomer materials, illumination causes photoisomerization of the dye. Due
to the shape change, the system loses equilibrium, the free energy increases
and anistoropic stress appears in the cell, which changes the shape of the
cell and of the bulk host, and can do mechanical work. Energy for the work
comes from the absorbed light. The model not only describes the mechanism
whereby particle shape change results in local increased free energy and
anisotropic internal stress, but allows quantitative estimates of these.

This paper has reported our first attempts to implement a simple mean field
theory for rod-like hard particles. Work to extend and improve the model is
clearly needed. Allowing the confined particles to respond to the
anisotropic internal pressures acting on them seems promising; as does
consideration of cell and particle shapes more general than those considered
here. Efforts in these directions are under way.

\paragraph{Acknowledgements}

P.P-M. acknowledges support from the Office of Naval Research through the
MURI on Photomechanical Material Systems (ONR N00014-18-1-2624). 

\end{document}